\documentclass[12pt]{iopart}
\usepackage{iopams} 
\usepackage{graphicx}
\usepackage[caption=false]{subfig}
\usepackage{color}
\usepackage{mathscinet}
\usepackage{epstopdf}
\pdfoutput=1
\begin{document}

\title[Reduction to Markovian dynamics]{Reduction from non-Markovian to
Markovian dynamics: The case of aging in the noisy-voter model}
\author{Antonio F. Peralta$^{1}$, Nagi Khalil$^{2}$, Ra\'ul Toral$^{1}$}
\address{(1) IFISC (CSIC-UIB), Instituto de F\'isica Interdisciplinar y Sistemas Complejos, Campus Universitat de les Illes Balears, E-07122 Palma de Mallorca, Spain.\\
(2) Escuela Superior de Ciencias Experimentales y Tecnolog{\'\i}a (ESCET),
Universidad Rey Juan Carlos, M\'ostoles 28933, Madrid, Spain.}

\begin{abstract}
We study memory dependent binary-state dynamics, focusing on the noisy-voter model. This is a non-Markovian process if we consider the set of binary states of the population as the description variables, or Markovian if we incorporate ``age", related to the time one has spent holding the same state, as a part of the description. We show that, in some cases, the model can be reduced to an effective Markovian process, where the age distribution of the population rapidly equilibrates to a quasi-steady state, while the global state of the system is out of equilibrium. This effective Markovian process shares the same phenomenology of the non-linear noisy-voter model and we establish a clear parallelism between these two extensions of the noisy-voter model. 
\end{abstract}

\date{\today}
\maketitle

\section{Introduction}\label{intro}

Agent-based models can be understood as a theoretical tool to analyze the mechanisms that are relevant and give rise to the collective behavior of a system composed by many interacting units. Typical examples of real phenomena that can be explained using agent-based models include, for example: the spatial distribution of electoral results~\cite{voter-election,Kononovicius2018,Kononovicius2019}, the time evolution of the number of speakers of a given language~\cite{Language1,Language4} and the evolution of prices in financial markets~\cite{Markets1,Lux2016}.

The stochastic description of these models is usually followed by a Markovian assumption. This implies that the rate at which different events in the system take place depend only on the present state of the system, i.e. memoryless. One of the most important properties of Markovian processes is that they exhibit an exponential distributions in the times for upcoming events and a Poissonian distribution in the number of events in a given time interval. Many real system, however, show strong discrepancies with this Markovian assumption. Empirical evidence of this can be observed, for example, when measuring inter-event time distributions of human activity, which happen to be heavy tailed~\cite{Candia,Karsai2012}, this is known in the literature as burstiness~\cite{Karsai2018}. The relaxation of the Markovian assumption has been considered and explored, from both theoretical and numerical points of views, in several contexts and models~\cite{Baxter2011,Takaguchi,Hoffmann2012,Min1,Boguna2014,Masuda2018,JEDRZEJEWSKI2018306}. Under certain circumstances and for specific models, a reduction of the non-Markovian to an effective Markovian process is possible~\cite{Gleeson,Feng2019}.

A well known technique to deal with non-Markovian processes is to extend the number of variables that describe the system to a level in which the system becomes Markovian~\cite{nonMarkov1,nonMarkov2}. Take a binary-state, ``spin'', model with $N$ individuals as an example with $k=1, ..., N$ being the label of the individuals and the set of states $\lbrace s_{k} \rbrace_{k=1}^{N}$ with $s_{k} = \pm 1$. If we incorporate to the set of states or spins the individual's internal age~\cite{perez2016competition}, also called persistence time (the time spent in a particular state), $\lbrace i_{k} \rbrace_{k=1}^{N}$ with $i_{k}=0, 1, 2,\dots$ (in arbitrary time units), we are including in our description a whole new variety of binary-state models. A Markovian model is defined by the rates, the probability per unit time, of individual $k$ changing its spin $\beta(s_{k} \rightarrow -s_{k})$, a rate which, in general, depends on the set of spins $\lbrace s_{k} \rbrace_{k=1}^{N}$ of the population. In a non-Markovian model, however, this rates depend additionally on the individual's age $\beta_{i_{k}}(s_{k} \rightarrow -s_{k})$. This generalization obviously does not refer to all the possible non-Markovian models that can be defined, but to a significant part.

As a prototypical example of opinion dynamics, we will study the noisy-voter (Kirman) model~\cite{Redner,kirman1993ants,Granovsky,Biancalani}. This model considers noisy/idiosyncratic changes of state and a copying/herding mechanism as the fundamental forces that drive the dynamics. Recent studies of the Markovian version of the model include: the effect of a network structure~\cite{Carro2,Peralta_pair}, external control of the system~\cite{Kononovicius,Carro1}, the role of zealots ~\cite{Nagi} and contrarians ~\cite{Nagi2}, more than two states~\cite{herrerias2019,Vazquez2019}, and first-passage properties~\cite{Nagi3}. Different non-Markovian versions have been studied as well for the model without noise ~\cite{Schweitzer,Juan,Oriol2,Peralta2019} and with noise~\cite{Oriol} (see~\cite{Artime2019} for a recent review of the model, both Markovian and non-Markovian versions). As a prominent real feature that the non-Markovian version of the model can reproduce, while the Markovian version can not, we should highlight the power law tails in the inter-event time distributions~\cite{Juan}.

In this paper, we focus on the mathematical methods to deal with such non-Markovian systems. We show that, in some cases, we can define {\slshape effective} rates $\beta(s_{k} \rightarrow -s_{k})$ that consider aging in an averaged way. This can be done when the age distribution of the population quickly reaches a quasi-steady state, where it only depends on the current set of spins $\lbrace s_{k} \rbrace_{k=1}^{N}$, and we can define thus an effective Markovian process with these rates. While the original rates $\beta_{i_{k}}(s_{k} \rightarrow -s_{k})$ of the noisy-voter model with aging are linear with respect to the spin variables $\lbrace s_{k} \rbrace_{k=1}^{N}$, the effective ones happen to be non-linear. We can expect then that some phenomenology of the noisy-voter model with aging could be equivalent to the one of the non-linear noisy-voter model. In fact, we find tristability and induced phase transitions, which are features of the non-linearity in the rates~\cite{Nyczka2012,Peralta,Jdrzejewski2019}, also in the aging version of the model~\cite{Oriol,Artime2019}.

The outline of the paper is as follows: in Section \ref{model_def} we define the ingredients of the noisy-voter model with and without aging and we review the main results in the existing literature. In  Section \ref{sec:master} we construct the general master equation of the noisy-voter model with aging and, in Section \ref{ada_sec} the steady state solutions of its deterministic dynamics are determined. In Section \ref{sec:adiabatic} we show that an adiabatic approximation of the stochastic dynamics is possible, exploring altogether its validity in the range of parameter values of the noise. In Section \ref{nonlinear} it is argued how this adiabatic elimination corresponds to a Markovian reduction of the dynamics, and the equivalence between the aging and non-linear versions of the model is discussed. We end with the conclusions and some discussion in Section \ref{sec_conc}. The technical details of the adiabatic elimination are explained in the \ref{app:attractor}.

\section{Model}
\label{model_def}

The standard noisy-voter model considers a system formed by $N$ individuals (or agents). They are located in the nodes of a single-connected network whose links define a neighborhood relationship: two individuals are neighbors of each other if they are connected by a link. Each individual $k=1,\dots,N$ holds a binary-state (spin) variable $s_k=\pm1$. The exact meaning of this binary-state variable depends on the interpretation of the model, e.g. the possible two languages A/B spoken by a bilingual person~\cite{Language1}, the susceptible/infected state with respect to an illness, the selling/buying state of a stock market broker~\cite{kirman1993ants}, the vote for republican/democrat in the USA election~\cite{voter-election}, etc., but its exact meaning does not concern us in this paper. The state variable can change over time by the following (stochastic) rules:\\[5pt]
(i) A node $k$ is selected at random amongst the $N$ possibilities.\\
(ii) With probability $a$ a new state $s_k=\pm1$ is chosen randomly.\\ 
(iii) Otherwise, hence with probability $1-a$, the individual copies the state $s_k=s_{k'}$ of another individual $k'$ chosen at random between the set of neighbors of $k$. \\[5pt]
Updates resulting of rule (ii) are called {\slshape noisy updates}, while those of rule (iii) reflect a {\slshape herding or imitation mechanism}. The parameter $a$ is called the noise intensity. The (noiseless) voter model takes $a=0$. Every time an individual is chosen for updating, time $t$ increases by one unit, while $N$ updates constitute one Monte Carlo step (MCS).

The important question one wants to address is whether, by repeated iteration of the above rules, the system reaches a situation of dominance of one of the two possible states, $s=\pm1$, or, on the contrary, each state is shared by approximately half of the total population. It turns out that the answer to this simple question is far from trivial.

In the case of the noiseless voter model, $a=0$, the full consensus states, those in which all individuals share the same state, are absorbing configurations as no further evolution is possible once one of those states is reached. In a finite system, due to the stochastic nature of the dynamics, the full consensus state is always reached by the noiseless voter model in a finite time. However, for effective spatial dimensions greater than two (this includes most common complex networks and the all-to-all connectivity considered later in this paper), the theoretical analysis, supported by numerical simulations, indicates that the system gets trapped in a dynamically active metastable state, in which the fraction of individuals holding the same state fluctuates around a constant value with fluctuations that decrease and tend to zero with increasing system size $N$. The system is able to scape this metastable state towards the full consensus state by means of a rare, large fluctuation whose likelihood decreases with $N$. The average time to scape the metastable state diverges with increasing $N$ and in the thermodynamic limit, $N\to\infty$, the full consensus states are never reached. The somehow counterintuitive result of the noiseless voter model is that by increasing the spatial dimension, i.e., by increasing the connectivity of the agents, the time to reach consensus increases significantly and, eventually, diverges in the thermodynamic limit, while one could naively have expected the opposite conclusion from the imitation mechanism, namely, that a poorly-connected set of individuals is less prone to achieve a situation of consensus while a well-connected society where everybody can interact with everybody else, would easily reach consensus.

For the noisy-voter model, $a>0$, the main difference is that the full consensus states are no longer absorbing configurations. The system becomes ergodic and it is able to reach any configuration starting from any other, so restoring the symmetry between the two possible states $s=\pm1$. If, using the Ising-like terminology, we define the {\slshape magnetization} $m=2x-1\in[-1,1]$ where $x$ is the fraction of agents holding the state $+1$, the stationary probability distribution $P(m)$ is symmetric around $m=0$ but its shape depends on the value of the noise term $a$ and the number of agents $N$. If the coefficient $a$ is greater that a critical value, $a>a_c(N)$, the noise term dominates and $P(m)$ has its maximum value at $m=0$, the coexistence state. For $a<a_c(N)$, the maxima of the distribution are at $m=\pm1$ (the full consensus states) and the most probable situation is one of consensus, although the consensus state fluctuates with time between the two values $s=\pm1$ restoring the symmetry of the two possible states. The exact expression for the critical value $a_c(N)$ depends on the details of the network, for instance for the all-to-all connectivity one finds $a_c(N)=2/(N+2)$, but it is a general feature of most networks that $a_c(N) \to 0$ in the thermodynamic limit $N\to\infty$. Therefore, in this limit, any finite noise term $a$ is always above the critical value $a_c=0$, and the most likely outcome of the model is that of coexistence of states, instead of the consensus found in the noiseless version. 

There have been several attempts to modify the model in order to make consensus possible or even the most likely outcome in the presence of noise. Some of them include a modification of the rule by which the probability of an agent to change its state by imitation depends on the absolute value, rather than on the fraction, of agents holding the opposite state~\cite{Markets2,alfarano2008time}, or by structuring the network of connectivities in a particular highly hierarchical star-like configuration~\cite{Markets3}. These seem rather arbitrary for most possible applications of the model. In this paper we consider the inclusion of a non-Markovian feature in the dynamics, namely the {\slshape internal age} or {\slshape inertia} of the agents, as an ingredient that allows (imperfect) consensus state to be the most probable outcome of the dynamics in some circumstances. 

The inertia of individuals affecting its willingness to change state by imitation was introduced in reference~\cite{Schweitzer} in the context of the noiseless voter model and it was later generalized in reference~\cite{Oriol} under the presence of noise. The basic idea is to introduce a mechanism, inertia or aging, by which agents are less prone to copy a neighbor's state the longer they have been holding their current state. The exact origin of the inertia again depends on the details of the interpretation of the model. In Physics, aging appears when the relaxation towards the stationary state displays slow dynamics and the time-translational invariance is broken. In the context of the binary-state models, it can refer to the increase to the resistance to an illness with increasing age (so making it more difficult to change the state from healthy to infected) or to the accommodativeness to a situation making it more difficult to change state. 

Within the specific setup of the voter model, we introduce the {\slshape internal age} $i_k=0,1,2,\dots$ of individual $k$ as a variable that stands for the number of update attempts elapsed since its last change of state. In the aging version of the noisy-voter model the above rules are modified such that the herding mechanism occurs only with an \emph{activation probability} $p_{i_k}$ that depends on the internal age $i_k$ of the selected individual. The evolution rules of the model are those of the noisy-voter model spelled out before modifying the last step as follows:\\[5pt]
(iii) Otherwise, hence with probability $1-a$, the individual copies {\bf with probability $\bf p_{i_k}$} the state $s_k=s_{k'}$ of another individual $k'$ chosen at random between the set of neighbors of $k$. If, either due to the noisy update or the herding mechanism, the selected individual $k$ has changed state $s_k \rightarrow -s_k$, then its internal age resets to zero $i_k \rightarrow 0$; otherwise it increases in one unit $i_k \rightarrow i_k+1$. Initially, all internal times are set to zero.

The standard noisy-voter model is recovered taking $p_i=1$ (or more generally, $p_i=p$, a constant setting the time scale of the process). Consistently with our definition of aging, $p_i$ should be a decreasing function of the age $i$. It turns out that for the noiseless voter model, the dynamical properties (e.g. whether or not the consensus state is reached asymptotically) depend on the detailed functional form of $p_i$, specifically on the way it tends to its limiting value $p_\infty$ and on whether $p_\infty$ is strictly zero or positive~\cite{Peralta2019}. 

In the context of the noisy-voter model, $a>0$, considered in this paper it turns out that those details are not so important and we present here results for a rational form for $p_i$, for which, due to its mathematical simplicity, one can perform most analytical calculations in full. Still, we have considered three qualitatively different scenarios depending on the specific functional form of $p_{i}$:
\begin{itemize}
\item[1.-] \emph{Aging:} In this case $p_{i}$ is a strictly decreasing function, thus $p_{i+1} < p_{i}$. As discussed before, this implies a reluctance to change state the longer an individual has been holding its current state. For the sake of concreteness and mathematical simplicity, we will be using the particular form 
\begin{equation}
\label{eq:paging}
p_i^{\mathrm{aging}}=\frac{b}{i+c},
\end{equation} 
where $c\ge b\ge0$ are constants. As mentioned before, the specific form in which the copying probability depends on the internal age is not really relevant for the noisy-voter model. Nevertheless, the particular rational dependence given by Eq.(\ref{eq:paging}) has been shown to induce features observed in several real-word systems, such as power-law inter-event time distributions~\cite{Juan}. Note that this rational form is a particular case of the more general case considered in~\cite{Peralta_voter} with $p_0=b/c$ and $p_\infty=0$.

\item[2.-] \emph{Anti-aging:} In this case $p_{i}$ is a strictly increasing function, $p_{i+1} > p_{i}$. As an individual spends more time in the same state, it is more likely to change state. A particular choice is
\begin{equation}
\label{eq:panti-aging}
p_i^{\mathrm{anti-aging}}=\frac{i+b}{i+c},
\end{equation}
where again $c\ge b>0$. This can be interpreted as individuals getting tired of their state and increasing the probability of copying another state the longer they have been holding their state. We note the particular values $p_0=b/c$ and $p_\infty=1$.

\item[3.-] \emph{Delayed aging:} In this case the aging mechanism only acts after a given internal age $i_0$: 
\begin{equation}
\label{eq:pdelayed}
p_i^{\mathrm{delayed\,aging}}=\cases{1,& for\quad $i< i_0$,\\p_{i-i_0}^{\rm{aging}},& {for\quad $i\ge i_0$.}}
\end{equation} This can be interpreted as a situation where young individuals behave very differently than older ones. Young individuals get bored easily and they are not able to hold the same opinion a long period of time, until they exceed a certain age and become less open to changes. 
\end{itemize}

As discussed in~\cite{Oriol}, the main result of the noisy-voter model in the presence of aging is that the system can sustain a consensus state up to a critical value of the noise, $a_c$, which has a non-zero value in the thermodynamic limit. For $a<a_c$, the majority of individuals share the same state and produce a non-null value of the magnetization $m$. In the thermodynamic limit, this partial consensus can occur in the state $s=+1$ or in $s=-1$ through a genuine symmetry-breaking second-order transition at $a=a_c$. For $a>a_c$ the magnetization $m$ fluctuates around zero with an amplitude that decreases with increasing $N$. Therefore, the absolute of the magnetization $|m|$ behaves as an order-parameter and displays critical behavior at $a=a_c$ that can be classified in the same class of universality as the Ising model. As aging impedes imitation, we conclude that, maybe counterintuitively, the inclusion of some reluctance to change favors the appearance of consensus.

In the case of anti-aging, not surprisingly, the main result~\cite{Ozaita} is that the system does not reach consensus for any value of the noise intensity $a$ and the magnetization $m$ fluctuates around zero with an amplitude that decreases with increasing $N$. More interesting is the case of delayed aging in which we will show that the transition to consensus at $a_c$ can become first-order with the existence of a tricritical point.

In the next sections we review in detail this phenomenology for the all-to-all connected network and obtain explicit expressions that allows us to determine the critical value $a_c$ as a function of the parameters $b,\,c,\,i_0$ of the model.

\section{The master equation}
\label{sec:master}
Throughout the remainder of this paper we use the all-to-all (or fully connected) network in which all nodes are neighbors. The effect of a more complex network structure in the interactions between nodes is a further complication in the mathematical description~\cite{Peralta_pair} and it is left for future studies. In the all-to-all setup, all the information needed to implement the stochastic update rules is contained in the set $S\equiv \lbrace n^{\pm}_{i} \rbrace_{i=0}^\infty$ of the numbers of individuals with internal age $i$ in states $\pm 1$, respectively~\cite{Schweitzer}. The global variables for the total up and down spins are $n = \sum_{i=0}^{\infty} n_{i}^{+}$ and $N-n= \sum_{i=0}^{\infty} n_{i}^{-}$. Observe that not all variables of the state $S$ are independent, since $\sum_{i=0}^\infty (n_i^++n_i^-)=N$. Hence, it is useful to consider an alternative representation of the system in terms of independent variables, for instance by using the variable $n$ and obviating $n_0^\pm$, as $S_n\equiv \left(n,\lbrace n_i^\pm\rbrace_{i=1}^\infty\right)$. In this representation $n_0^\pm$ are given by $n_0^{+}=n-\sum_{i=1}^\infty n_i^+$ and $n_0^{-}=N-n-\sum_{i=1}^\infty n_i^-$.

The stochastic update rules of the model induce four independent processes that modify the values of the set $S$ and whose respective probabilities can be computed as follows~\cite{Oriol,Peralta_voter}:

{\bf(1)} Consider that at time $t$ the chosen individual $k$ is in state $s_k=+1$, has age $i_k=i$ and it switches to $s_k=-1$ such that its age is reset to $i_k=0$. The occurrence of this process demands first choosing, with probability $\frac{n_i^+}{N}$, an individual with age $i$ and state $+$. Then this individual, with probability $a$ can undergo a noisy update and choose the state $i_k=-1$ with probability $1/2$. Alternatively, with probability $1-a$, the herding mechanism acts with a probability that results of multiplying the age-dependent probability, $p_i$, by the probability $\frac{N-n}{N}$ that the randomly selected neighbor is in the opposite state. Altogether, the probability is $\frac{n_i^+}{N} \left[\frac{a}{2}+(1-a)p_i \frac{N-n}{N}\right]\equiv \frac{1}{N}\Omega_{1,i}$. When the switching of state occurs, we have $n_{i}^{+} \rightarrow n_{i}^{+}-1$ and $n_0^{-} \rightarrow n_0^{-}+1$.\\

{\bf(2)} This is similar to the previous case but now the chosen individual is initially in state $s_k=-1$. The probability of switching is $\frac{n_i^-}{N} \left[\frac{a}{2}+(1-a)p_i \frac{n}{N}\right]\equiv \frac{1}{N}\Omega_{2,i}$. When the switching of state occurs, we have $n_{i}^{-} \rightarrow n_{i}^{-}-1$ and $n_0^{+} \rightarrow n_0^{+}+1$.\\

{\bf(3)} Consider that at time $t$ the chosen individual $k$ has $s_k=+1$ and age $i_k=i$, but that now it keeps its current state $s_k=+1$. This event happens with a probability equal to the probability of choosing an individual in state $+1$ with age $i$, $\frac{n_i^+}{N}$, multiplied by the probability that it does not switch, which can arise either because the noisy update, of probability $a$, generated, with probability $1/2$, the same state $s_k=+1$ or, alternatively, with probability $1-a$, either the copying mechanism was not activated, with probability $1-p_i$, or it was activated, probability $p_i$, but the selected neighbor was also in the state $+1$, with probability $\frac{n}{N}$. Altogether, the probability is $\frac{n_i^+}{N} \left[\frac{a}{2}+(1-a)\left(1-p_i+p_{i}\frac{n}{N} \right)\right] \equiv \frac{1}{N}\Omega_{3,i}$. In this case the variables change as $n_{i}^{+} \rightarrow n_{i}^{+}-1$ and $n_{i+1}^{+} \rightarrow n_{i+1}^{+}+1$.\\

{\bf(4)} Finally, we consider a similar case to the previous one but the chosen individual $s_k=-1$ keeps its state. The switching probability is now $\frac{n_i^-}{N}\left[\frac{a}{2}+(1-a)\left(1-p_i+p_i\frac{N-n}{N}\right)\right]\equiv\frac{1}{N}\Omega_{4,i}$. The changes in the state $S$ are $n_i^-\to n_i^--1$, $n_{i+1}^-\to n_{i+1}^-+1$.\\

The derivation has considered a discrete-time process in which time increases by $dt=1/N$ after each trial. We can consider instead a continuous-time process whose rates are obtained by dividing the corresponding probabilities by the time step $dt$:
\begin{eqnarray}
\label{rates1}
n^{+}_{i}&\rightarrow& n^{+}_{i} - 1, \hspace{0.5cm} n^{-}_{0} \rightarrow n^{-}_{0} + 1: \hspace{1.0cm} \Omega_{1,i}(S) = n^{+}_{i} \beta_{i}(1-x), \\
\label{rates2}
n^{-}_{i}&\rightarrow& n^{-}_{i} - 1, \hspace{0.5cm} n^{+}_{0} \rightarrow n^{+}_{0} + 1: \hspace{1.0cm} \Omega_{2,i}(S) = n^{-}_{i} \beta_{i}(x), \\
\label{rates3}
n^{+}_{i}&\rightarrow& n^{+}_{i} - 1, \hspace{0.5cm} n^{+}_{i+1} \rightarrow n^{+}_{i+1} + 1: \hspace{0.5cm} \Omega_{3,i}(S)= n^{+}_{i} \alpha_{i}(1-x), \\
\label{rates4}
n^{-}_{i}&\rightarrow& n^{-}_{i} - 1, \hspace{0.5cm} n^{-}_{i+1} \rightarrow n^{-}_{i+1} + 1: \hspace{0.5cm} \Omega_{4,i}(S) = n^{-}_{i} \alpha_{i}(x),
\end{eqnarray}
where $x=n/N$ and we have defined the functions
\begin{eqnarray}
\label{alpha}
\alpha_{i}(x) = \frac{a}{2}+(1-a) \left(1 - p_{i} x \right), \hspace{0.5cm} \beta_{i}(x)=1-\alpha_i(x)= \frac{a}{2}+(1-a) p_{i} x.
\end{eqnarray}
In this context, $\beta_i(1-x)$ is the rate that an individual of age $i$ in state $+1$ changes state to $-1$ and similar interpretations for $\beta_i(x)$, $\alpha_i(x)$ and $\alpha_i(1-x)$. This model is said to be {\slshape linear} because the functions $\alpha_i$ and $\beta_i$ depend linearly on the fraction $x=n/N$ or $1-x=(N-n)/N$ of individuals in the opposite state. 

By means of an standard probabilistic balance, and in the continuous time limit, we can construct the master equation~\cite{vKampen, Peralta_moments} for the probability $P(S; t)$ of the state of the system being $S=\lbrace n^{\pm}_{i} \rbrace_{i=0}^\infty$ at time $t$ (measured in MCS):
\begin{eqnarray}
\label{Master_equation}
 \frac{\partial P(S;t)}{\partial t} &=& \sum_{i=0}^{\infty} \left[ \left( E_{n_{i}^{+}}^{+} E_{n_{0}^{-}}^{-} - 1 \right) [\Omega_{1,i} P] + \left( E_{n_{i}^{-}}^{+} E_{n_{0}^{+}}^{-} - 1 \right) [\Omega_{2,i} P] \right. \nonumber\\
&&+ \left. \left( E_{n_{i}^{+}}^{+} E_{n_{i+1}^{+}}^{-} - 1 \right) [\Omega_{3,i} P] + \left( E_{n_{i}^{-}}^{+} E_{n_{i+1}^{-}}^{-} - 1 \right) [\Omega_{4,i} P] \right],
\end{eqnarray}
where $E_{n_{i}^{+}}^{\pm}$ are step operators defined as $E_{n_{i}^{+}}^{\pm} f(\dots,n_{i}^{+},\dots)=f(\dots,n_{i}^{+}\pm 1,\dots)$ and $E_{n_{i}^{-}}^{\pm} f(\dots,n_{i}^{-},\dots)=f(\dots,n_{i}^{-}\pm 1,\dots)$ for any function $f$ defined on the state $S$.

We also mention the master equation satisfied in the $S_n$ representation, obtained taking into account explicitly the effect of the $i=0$ terms of Eq.(\ref{Master_equation}) on the variables $\left(n,\lbrace n_i^\pm\rbrace_{i=1}^\infty\right)$:
\begin{eqnarray}
\label{Master_equation_sx}
 \frac{\partial P(S_n;t)}{\partial t} &=& \sum_{i=1}^{\infty} \left[ \left( E_{n_{i}^{+}}^{+} E_{n}^{+} - 1 \right) [\Omega_{1,i} P] + \left( E_{n_{i}^{-}}^{+} E_{n}^{-} - 1 \right) [\Omega_{2,i} P] \right. \nonumber\\
&&+ \left. \left( E_{n_{i}^{+}}^{+} E_{n_{i+1}^{+}}^{-} - 1 \right) [\Omega_{3,i} P ]+ \left( E_{n_{i}^{-}}^{+} E_{n_{i+1}^{-}}^{-} - 1 \right) [\Omega_{4,i} P ] \right] \nonumber\\
&& + \left( E_{n}^{+} - 1 \right)[\Omega_{1,0} P]+ \left( E_{n}^{-} - 1 \right)[ \Omega_{2,0} P] \nonumber\\
&& + \left( E_{n_{1}^{+}}^{-} - 1 \right)[ \Omega_{3,0} P]+ \left( E_{n_{1}^{-}}^{-} - 1 \right) [\Omega_{4,0} P].
\end{eqnarray}
where $E_n^\pm f(n,n_1^\pm,n_2^\pm,\dots)=f(n\pm 1,n_1^\pm,n_2^\pm,\dots)$ for any function $f$ defined on the state $S_n$. Here, the rates $\Omega_{\ell,0}$ must be written in terms solely of the variables of $S_n$ using the functional relation between $n_0^\pm$ and $\{n,n_i^\pm\}_{i=1}^\infty$. In the next section we obtain the steady-state average values of this stochastic process.

\section{Mean-field analysis: steady states}\label{ada_sec}

From the master equation (\ref{Master_equation_sx}) we can obtain the evolution equations for the ensemble average of the fraction of nodes with a given state and age $x_i^\pm=\langle n_i^\pm\rangle/N$, as well as for $x = \langle n \rangle/N$. We use the mean-field approximation which neglects correlations as $\langle n_i^\pm n \rangle \simeq \langle n_i^\pm\rangle \langle n \rangle$ and end up with a closed infinite system of equations:
\begin{eqnarray}
\label{eq_dyn1}
\frac{d x_{i}^{+}}{dt} &=& -x_{i}^{+} + x_{i-1}^{+} \alpha_{i-1}(1-x), \quad i\ge 1, \\
\label{eq_dyn2}
\frac{d x_{i}^{-}}{dt} &=& -x_{i}^{-} + x_{i-1}^{-} \alpha_{i-1}(x), \quad i\ge 1,\\
\label{eq_dyn3}
\frac{dx}{dt} &=& \sum_{i=0}^{\infty} x_{i}^{-} \beta_{i}(x) - \sum_{i=0}^{\infty} x_{i}^{+} \beta_{i}(1-x).
\end{eqnarray}
In these equations, variables $x_0^\pm$, whenever they appear, should be expressed in terms of the independent variables,
\begin{eqnarray}
 \label{eq:x01}
 x_0^+=x-\sum_{i=1}^{\infty}x_{i}^{+}, \qquad x_0^- = 1-x - \sum_{i=1}^{\infty} x_{i}^{-}.
\end{eqnarray}
The explicit time evolution of these variables is
\begin{eqnarray}
\label{eq_dyn4}
\frac{d x_{0}^{+}}{dt} &=& -x_{0}^{+} + \sum_{i=0}^{\infty} x_{i}^{-} \beta_{i}(x), \\
\label{eq_dyn5}
\frac{d x_{0}^{-}}{dt} &=& -x_{0}^{-} + \sum_{i=0}^{\infty} x_{i}^{+} \beta_{i}(1-x).
\end{eqnarray}

By equating all time derivatives to zero, we can identify the steady-state solutions for the mean-field description. From Eqs.~(\ref{eq_dyn1},\ref{eq_dyn2},\ref{eq:x01}) we find
\begin{eqnarray}
\label{eq:xi1}
&& x_{0,\mathrm{st}}^+=\frac{x_\mathrm{st}}{f(x_\mathrm{st})}, \hspace{3.1cm} x_{0,\mathrm{st}}^-=\frac{1-x_\mathrm{st}}{f(1-x_\mathrm{st})},\\
 \label{eq:x03}
&& x_{i,\mathrm{st}}^+=\frac{x_\mathrm{st}}{f(x_\mathrm{st})}\prod_{k=0}^{i-1}\alpha_{k}(1-x_\mathrm{st}), \hspace{0.5cm} x_{i,\mathrm{st}}^-=\frac{1-x_\mathrm{st}}{f(1-x_\mathrm{st})}\prod_{k=0}^{i-1}\alpha_{k}(x_\mathrm{st}),\quad i\ge 1,
\end{eqnarray}
where\footnote{In~\cite{Oriol} we used the notation $f(a,x)$ to stress the dependence on $a$ of this function.}.
\begin{equation}
\label{eq:funf}
 f(x)\equiv 1+\sum_{i=1}^\infty \prod_{j=0}^{i-1}\alpha_{j}(1-x).
\end{equation}
An obvious difficulty appears when the series Eq.(\ref{eq:funf}) defining $f(x)$ is not convergent, but this is never the case for $a>0$ (the only one considered in this paper), since $\alpha_j\le 1-a/2<1$ and it follows $f(x) \le \frac{2}{a}$. We realize that possible difficulties may arise for the case $a=0$, specially for $p_{i\rightarrow \infty}\rightarrow 0$, when d'Alembert's criterion does not ensure convergence as $\lim_{i\rightarrow \infty} \alpha_{i}(x) = 1$, see~\cite{Peralta_voter}.

Using Eqs.(\ref{eq_dyn3},\ref{eq_dyn4},\ref{eq_dyn5}) in the steady state, one obtains easily $x_{0,\mathrm{st}}^+=x_{0,\mathrm{st}}^-$, or
\begin{equation}
 \label{eq:xest}
 \frac{x_{\mathrm{st}}}{f(x_{\mathrm{st}})}=\frac{1-x_{\mathrm{st}}}{f(1-x_{\mathrm{st}})}.
\end{equation}
The solutions to this equation provide the possible steady-state values of $x_\mathrm{st}=\langle n\rangle /N$ and those values of $x_\mathrm{st}$ determine the other quantities, through Eqs.~(\ref{eq:xi1},\ref{eq:x03}). It is clear that $x_\mathrm{st}=1/2$ is always a trivial solution that corresponds to a symmetric steady state, with the same mean number of nodes with a given age having opposite states $x_{i,\mathrm{st}}^+=x_{i,\mathrm{st}}^-$ for $i\ge 0$. Other non-trivial solutions $x_\mathrm{st}(a)$ might appear depending on the function $f(x)$ and its dependence with the parameters of the system, e.g. the noise intensity $a$. Note that if $x_\mathrm{st}(a)$ is a solution, then $1-x_\mathrm{st}(a)$ is a solution as well. In any case, the steady-state solutions describe situations where $x_{i,\mathrm{st}}^\pm$ are decreasing functions of the age. 

Let us now give explicit expressions for $f(x)$ for the three possible cases introduced before.
\begin{itemize}
\item[1.-] \emph{Aging:} General analytical expressions can be obtained for arbitrary $b$ and $c$ in terms of hypergeometric functions\footnote{The expression is $f^{\mathrm{aging}}(x)=1+\left(1-\frac{a}{2}\right)v(x){_2F_1}\left(1,1+cv(x),1+c,1-\frac{a}{2}\right)$ with $v(x)\equiv1-\frac{1-a}{1-a/2}\frac{b}{c}(1-x)$.} but we reproduce here the simple expressions for the particular case $b=1,\,c=1$:
\begin{eqnarray}
\label{f_aging}
f^{\mathrm{aging}}(x) = \left( 2/a \right)^{1-\kappa(1-x)},\quad \kappa\equiv\frac{1-a}{1-a/2} 
\end{eqnarray}

\item[2.-] \emph{Anti-aging:} Again, general analytical expressions can be obtained for arbitrary $b$ and $c$ in terms of hypergeometric functions, but we reproduce here the simple expression valid for the case $b=0$, $c=1$:
\begin{equation}
\label{f_anti}
f^{\mathrm{anti-aging}}(x) = \left[1-g(x)\right]^{\frac{-1}{1-\kappa (1-x)}},
\end{equation}
with $g(x)=\frac{a}{2}+(1-a)x$ and $\kappa$ as in Eq.(\ref{f_aging}).

\item[3.-] \emph{Delayed aging:} A detailed calculation leads to the general expression:
\begin{equation}
\label{f_mixed2}
f^{\mathrm{delayed\,aging}}(x) = \frac{1-[g(1-x)]^{1+i_{0}}}{1-g(1-x)} + [g(1-x)]^{i_{0}}f^{\mathrm{aging}}(x).
\end{equation}
\end{itemize}

It is now possible to find the solutions to Eq.(\ref{eq:xest}) and obtain the steady states $x_{\mathrm{st}}$ as a function of the parameters of the model. Except in the case of anti-aging for which the only solution is $x_{\mathrm{st}}=1/2$, several solutions are possible and we need to establish their stability in order to derive the phase diagram. We postpone the discussion to Section \ref{sub:phasediagram} where we write the stationary distribution function in terms of a potential function whose absolute minima determine the stable phases.

\section{Adiabatic elimination}
\label{sec:adiabatic}
In the previous section we have been able to find the steady state value of the average number of nodes $n$ in the $+$ state using as a starting point a description in terms of the infinite set of variables $S_n=\left(n,\lbrace n_i^\pm\rbrace_{i=1}^\infty\right)$. The necessity of such a complicated description arises, obviously, from the non-Markovian nature of the aging process. However, due to the mathematical difficulties, it does not seem to be possible to derive from this detailed description other properties such as the time evolution of $\langle n(t)\rangle$ nor its fluctuations $\sigma^2_{n}(t) = \langle n(t)^2\rangle-\langle n(t)\rangle^2$. 

It is indeed hypothetically possible, at the mean-field level, to obtain a closed evolution equation for the variable $x(t)=\langle n(t)\rangle/N$. However, this equation depends on the whole range of previous states of the variable $x(s \leq t)$, as it is characteristic of non-Markovian processes. The equation can be obtained integrating first Eqs.(\ref{eq_dyn1}, \ref{eq_dyn2}), which leads to integral expressions for $x_{i \geq 1}^{\pm} (t)$ as a function of $x_{0}^{\pm}(s \leq t)$ and $x(s \leq t)$. Introducing this in Eq.(\ref{eq_dyn3}) we obtain the time evolution of $x(t)$ as function of $x(s \leq t)$ and $x_{0}^{\pm}(s \leq t)$. After eliminating, if possible, $x_{0}^{\pm}(s \leq t)$ in this equation using the constraints Eq.(\ref{eq:x01}), we find an integro-differential equation for $x(t)$. This can be done in detail for the noiseless $a=0$ case~\cite{Peralta_voter}, but an extension to $a>0$ seems to be impeded again by mathematical difficulties.

Our aim in this section is to derive an approximate closed description of the stochastic process in terms of the global variable $n$. To this end we use an adiabatic approximation in which $n(t)$ is considered to be a slow variable to which the other variables $\lbrace n_i^\pm(t)\rbrace_{i=1}^\infty$ are enslaved to. The problem of adiabatic elimination of fast variables in stochastic processes has been considered in a large class of problems~\cite{SanMiguel:1982,Oppo:1986,Brion:2007,pineda:2009}. We follow here closely the approach by Haken~\cite{Haken}. First we present the derivation of a closed master equation for the stochastic variable $n$ and next we justify the use of the adiabatic elimination.

\subsection{Derivation of a closed master equation for $n$}\label{sec_adiabatic}
We split the probability as 
\begin{equation}
P(S_n;t)=H(\{n_{i}^{\pm}\}_{i=1}^\infty;t \vert n) G(n;t),
\end{equation}
with $H(\{n_{i}^{\pm}\}_{i=1}^\infty;t \vert n)$ the conditional probability of the set $\{n_i^\pm\}_{i=1}^\infty$ to a value of $n$, and $G(n;t)$ the probability function of the global variable $n$. Inserting this in the master equation (\ref{Master_equation_sx}) and summing over $\{n_{i}^{\pm}\}_{i=1}^\infty$ we find
\begin{equation}
\label{master-G}
\frac{\partial G}{\partial t} = \left( E_{n}^{+} - 1 \right) [\widehat{\Omega}_{1}G]+ \left(E_{n}^{-} - 1 \right) [\widehat{\Omega}_{2} G], 
\end{equation}
with rates 
\begin{equation}
 \widehat{\Omega}_{\ell}=\sum_{j=1}^\infty\sum_{\{n_{i}^{\pm}\}_{i=1}^\infty} \Omega_{\ell,j}(S_n)H(\{n_{i}^{\pm}\}_{i=1}^\infty,t \vert n), \qquad \ell=1,2.
\end{equation}
The equation for the conditional probability $H(\{n_{i}^{\pm}\}_{i=1}^\infty;t \vert n)$ can be obtained from Eq.(\ref{Master_equation_sx}) using an adiabatic approximation that assumes that variables $\{n_{i}^{\pm}\}_{i=1}^{\infty}$ are enslaved to the evolution dictated by $n(t)$. In practice~\cite{Haken}, this means to consider that transitions between the $\{n_{i}^{\pm}\}_{i=1}^\infty$ variables occur for fixed $n$. This is tantamount to replacing $E^{\pm}_n$ by the identity operator. The resulting equation for $H$ is not written out in full, but we note that the equations for the time evolution of the conditioned mean values $\langle n_i^\pm|n\rangle$ are identical with Eqs.(\ref{eq_dyn1},\ref{eq_dyn2}) fixing the value of $x=n/N$.

According to this analysis , the evolution of the system takes place in two stages. A first one where the probability $H(\{n_{i}^{\pm}\}_{i=1}^\infty,t \vert n)$ of a given set of values $\{n_{i}^{\pm}\}$ conditioned to a value of $n$ rapidly evolves to its steady-state form, and a second stage where the probability of the set $\{n_{i}^{\pm}\}$ is slaved to a value of $n$. That is, after a short transient, the dynamics is completely given by the time evolution of the global variable $n$. Hence, at the last stage of the dynamics, we can consider the dynamics of $n(t)$ with transitions $n \rightarrow n \pm 1$ and respective rates $\widehat{\Omega}_{1,2}$ of Eq.(\ref{master-G}) after replacing $n_{i}^{\pm}$ by the stationary average value $\langle n_{i}^{\pm} \vert n \rangle_{\mathrm{st}}$. Note that the rates Eqs.(\ref{rates1},\ref{rates2}) are linear with $n_{i}^{\pm}$ conditioned to $n$ and we only need the average value $\langle n_i^\pm|n\rangle$ to compute the rates $\widehat{\Omega}_{1,2}$. As we mentioned before, the equations for the time evolution of $\langle n_{i}^{\pm} \vert n \rangle$ are Eqs.(\ref{eq_dyn1},\ref{eq_dyn2}) with $x=n/N$ regarded as a parameter. The stationary average values are then equivalent to Eqs.(\ref{eq:xi1},\ref{eq:x03}) in its extensive version, this is:
\begin{eqnarray}
\label{average_cond1}
\langle n_{i}^{+} \vert n \rangle_{\mathrm{st}} &=& \frac{n}{f(x)} \prod_{k=0}^{i-1}\alpha_{k}(1-x), \hspace{1.0cm} i \geq 1, \\
\label{average_cond2}
\langle n_{i}^{-} \vert n \rangle_{\mathrm{st}} &=& \frac{N-n}{f(1-x)} \prod_{k=0}^{i-1}\alpha_{k}(x), \hspace{1.0cm} i \geq 1, 
\end{eqnarray}
with $\langle n_{0}^{+} \vert n \rangle_{\mathrm{st}} = n/f(x)$ and $\langle n_{0}^{-} \vert n \rangle_{\mathrm{st}} = (N-n)/f(1-x)$. Thus we can compute the rates as:
\begin{eqnarray}
\label{ratesp}
\widehat{\Omega}_1 &=& \sum_{i=0}^{\infty} \Big[\Omega_{1,i} \Big]_{n_{i}^{+} = \langle n_{i}^{+} \vert n \rangle_{\mathrm{st}}} = \frac{n}{f(x)},\\
\label{ratesm}
\widehat{\Omega}_2&=& \sum_{i=0}^{\infty} \Big[\Omega_{2,i} \Big]_{n_{i}^{-} = \langle n_{i}^{-} \vert n \rangle_{\mathrm{st}}} = \frac{N-n}{f(1-x)},
\end{eqnarray} 
where we have used the property $\beta_{0}+\sum_{i=1}^{\infty} \beta_{i} \prod_{k=0}^{i-1} \alpha_{k}=1$, as $\beta_{i}=1-\alpha_{i}$. 

In Section \ref{nonlinear} we will analyze the predictions of the master equation (\ref{master-G}) with the above rates. Now we elaborate on the validity of the adiabatic approximation.

\subsection{Justification of the validity of the adiabatic elimination}

In order to justify the adiabatic approximation, we perform in this section an analysis based on the dynamics of the mean field description Eqs.(\ref{eq_dyn1}-\ref{eq_dyn3}). First of all, we identify a special solution of Eqs.(\ref{eq_dyn1}-\ref{eq_dyn3}) where all time dependence of $\{x_i^\pm\}$ occurs through $x(t)$. This solution, which is labelled with a subindex $s$, will be identified as slow and to be an attractor, or center manifold, of the dynamics. Introducing this proposed solution $x^{\pm}_{i,s}(t)=x^{\pm}_{i,s}(x(t))$ in Eqs.(\ref{eq_dyn1}-\ref{eq_dyn3}) we find:
\begin{eqnarray}
\frac{d x_{i,s}^{+}}{dx} &=& \frac{-x_{i,s}^{+} + x_{i-1,s}^{+} \alpha_{i-1}(1-x)}{\sum_{i=0}^{\infty} x_{i,s}^{-} \beta_{i}(x) - \sum_{i=0}^{\infty} x_{i,s}^{+} \beta_{i}(1-x)}, \quad i\ge 1, \label{atractor1} \\
\frac{d x_{i,s}^{-}}{dx} &=& \frac{-x_{i,s}^{-} + x_{i-1,s}^{-} \alpha_{i-1}(x)}{\sum_{i=0}^{\infty} x_{i,s}^{-} \beta_{i}(x) - \sum_{i=0}^{\infty} x_{i,s}^{+} \beta_{i,s}(1-x)}, \quad i\ge 1, \label{atractor2}
\end{eqnarray}
where $x_{0,s}^\pm$ should be written in terms of $x^{\pm}_{i,s}$ with $i\ge 1$ and $x$. As it becomes apparent, Eqs.(\ref{atractor1},\ref{atractor2}) are not easy to solve. They even represent a challenge from the numerical point of view, as the solution must satisfy the boundary conditions $x_{i,s}^+(x=0)=x_{i,s}^-(x=1)=0$. Note also the additional property $x_{i,s}^+(x)=x_{i,s}^-(1-x)$.

The evolution of $x(t)$ within the attractor is then obtained by solving the equation resulting from using the solution of the latter system $x^{\pm}_{i,s}(x)$ in Eq.(\ref{eq_dyn3}), this is:
\begin{equation}
\label{atractor3}
 \frac{dx}{dt} = \sum_{i=0}^{\infty} x_{i,s}^{-}(x) \beta_{i}(x) - \sum_{i=0}^{\infty} x_{i,s}^{+}(x) \beta_{i}(1-x). 
\end{equation}
The exact trajectory $x^{\pm}_{i,s}(x)$ and its rigorous analysis is based on the centre manifold theory~\cite{Oppo:1986,Lugiato,Ariel}, which in this context of Eqs.(\ref{atractor1},\ref{atractor2},\ref{atractor3}) is too complicated to carry out. In \ref{app:attractor} we calculate the first terms of the expansion of the attractor around a fixed point to illustrate the difficulties found. A crude simplification is the bare adiabatic elimination which assumes that $x_{i,s}^{\pm}(x)$ are determined by setting the numerator of Eqs.(\ref{atractor1},\ref{atractor2}) to zero, this is:
\begin{equation}
\label{eq:or0}
 0\simeq -x_{i,s}^++x_{i-1,s}^+\alpha_{i-1}(1-x); \quad 0\simeq -x_{i,s}^-+x_{i-1,s}^-\alpha_{i-1}(x); \quad i\ge 1,
\end{equation}
whose solution reproduces Eqs.(\ref{eq:xi1},\ref{eq:x03}), namely
\begin{eqnarray}
 \label{eq:at03}
x_{i,s}^+(x)=\frac{x}{f(x)}\prod_{k=0}^{i-1}\alpha_{k}(1-x), \hspace{0.5cm} x_{i,s}^-(x)=\frac{1-x}{f(1-x)}\prod_{k=0}^{i-1}\alpha_{k}(x),\quad i\ge 0,
\end{eqnarray}
with $f(x)$ given by Eq.(\ref{eq:funf}) and the convention $\prod_{k=0}^{-1}\alpha_{k}\equiv 1$. Note that this approximate solution does satisfy the correct conditions $x_{i,s}^+(x=0)=x_{i,s}^-(x=1)=0$, $x_{i,s}^+(x)=x_{i,s}^-(1-x)$. Furthermore, it predicts $x_{i,s}^+(x=1)=x_{i,s}^-(x=0)=\frac{a}{2}\left(1-\frac{a}{2}\right)^i$, i.e. a geometric distribution.
 
Replacing Eqs.(\ref{eq:at03}) in Eq.(\ref{atractor3}), and using the property $\alpha_k(x)+\beta_k(x)=1$, we obtain:
\begin{equation}
\label{eq:x}
 \frac{dx}{dt}=F(x)= {\cal C} \left[ \frac{1-x}{f(1-x)}-\frac{x}{f(x)} \right],
\end{equation}
with ${\cal C} =1$. The derivation of this simple, explicit, form for the attractor is probably not completely satisfactory. Nevertheless, it is possible to improve this approximate equation by imposing that the first two coefficients of the expansion of $F(x)$ around the fixed point $\tilde x=1/2$, $F(x)=\varepsilon_1(x-\tilde x)+\varepsilon_3(x-\tilde x)^3$, coincide with the first two coefficients of the exact expansion of the attractor around the same point, as computed in \ref{app:attractor}. It turns out that this can be achieved by including a constant ${\cal C}$ of order $1$ in the definition of $F(x)$ in Eq.(\ref{eq:x}). \ref{app:attractor} provides an explicit expression for this multiplicative constant. For example, for the aging probability $p_i=1/(i+2)$ it is ${\cal C}=0.316\dots$

\begin{figure}[t!]
\centering
\subfloat[]{\label{fig:tau:a}\includegraphics[width=0.45\textwidth]{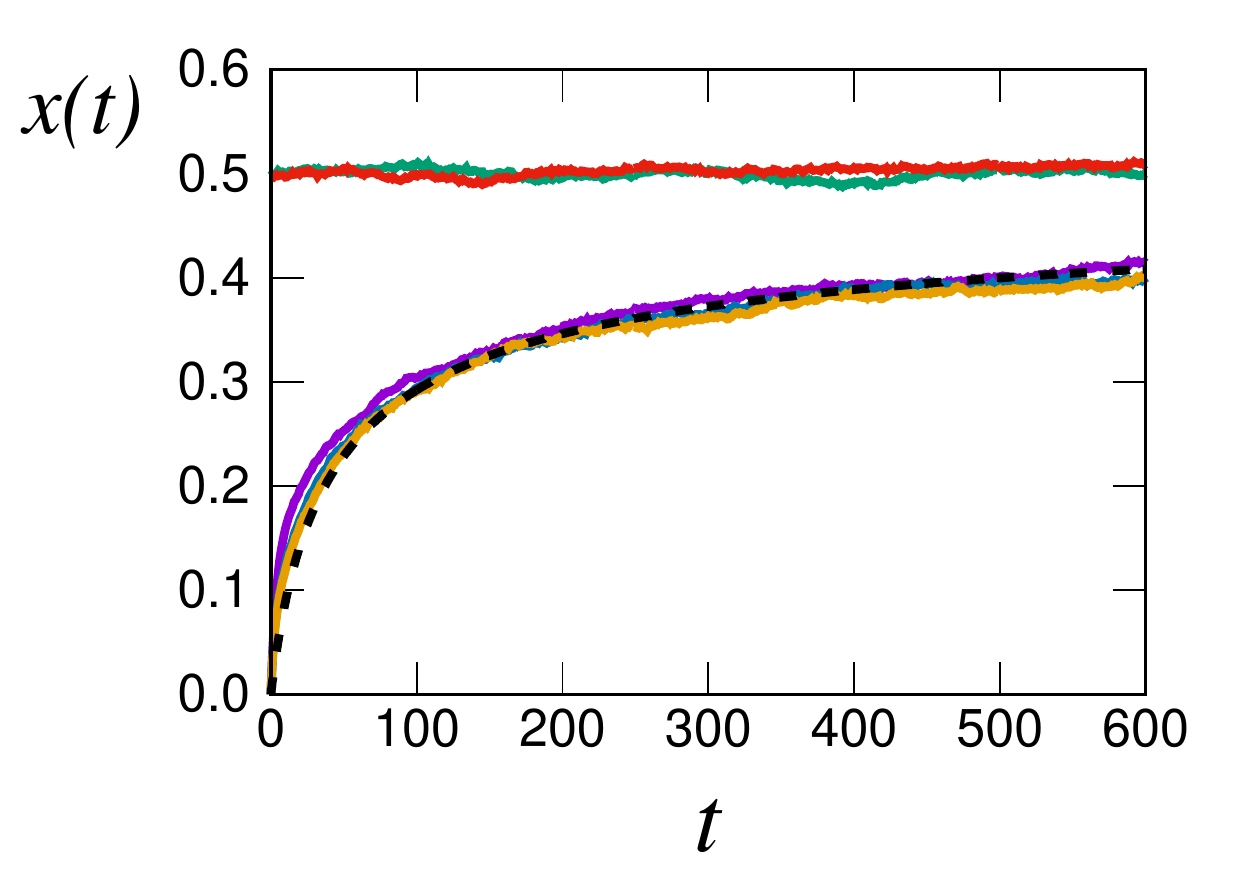}}
\subfloat[]{\label{fig:tau:b}\includegraphics[width=0.45\textwidth]{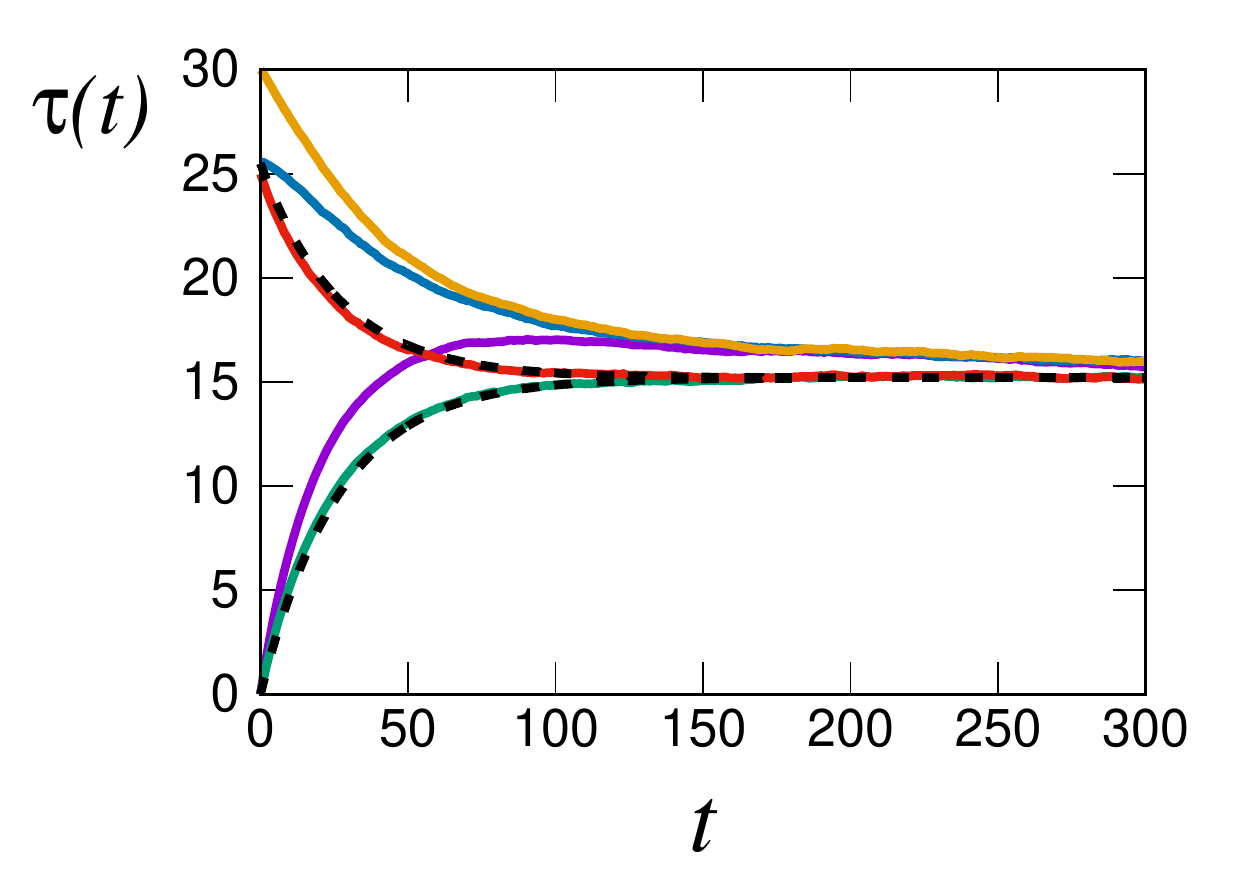}}

\subfloat[]{\label{fig:tau:c}\includegraphics[width=0.45\textwidth]{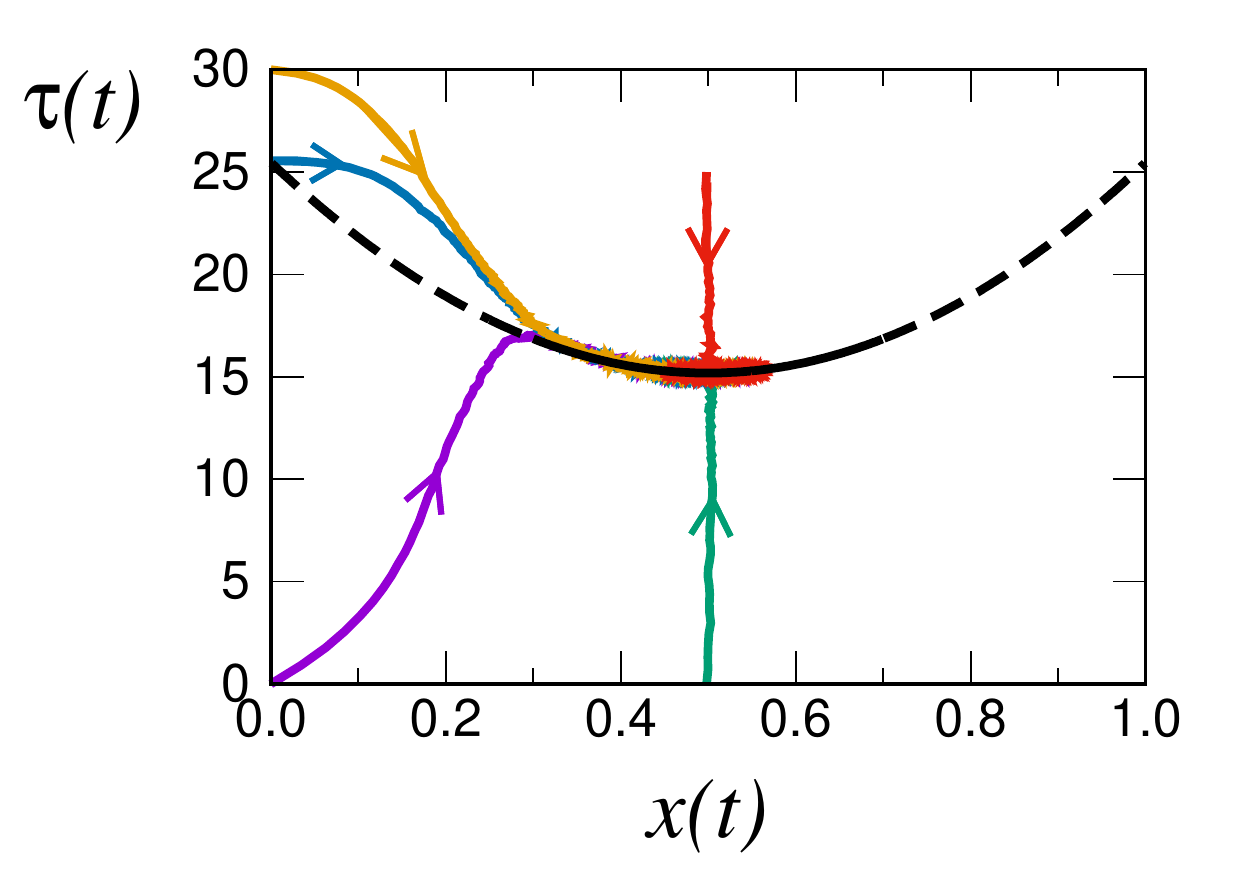}}
\caption{Time evolution of the global variable $x(t)$ and mean internal time $\tau(t)$, as defined in the main text. Color lines are trajectories coming from direct numerical simulations of the process given by  Eqs.(\ref{rates1}-\ref{rates4}) using $a = a_{c} \simeq 0.07556$, $p_{i}=(2+i)^{-1}$ and $N=10^5$, and the following initial conditions: (red) $x_i^+(0)=x_i^-(0)=\frac12\delta_{i,25}$; (green) $x_i^+(0)=x_i^-(0)=\frac12\delta_{i,0}$; (blue) the geometric distribution $x_{i}^-(0)=\frac{a}{2}\left(1-\frac{a}{2}\right)^i$, $\,x_i^+(0)=0$; (purple) $x_i^-(0)=\delta_{i,0}$, $x_i^+(0)=0$; (yellow) $x_i^-(0)=\delta_{i,30}$, $x_i^+(0)=0$. The dashed black line of panel (a) corresponds to the solution $x(t)$ of Eq.(\ref{eq:x}) with ${\cal C} = 0.316...$ and $x(0)=0$. The dashed black lines of panel (b) follow the functional form $\tau(t)=\tau_{s}(x=1/2)+(\tau(0)-\tau_{s}(x=1/2))e^{-at/2}$ with the two initial conditions $\tau(0)=0$ (matching the red curve) and $\tau(0)=25$ (matching the green curve). The black solid line of panel (c) corresponds to the approximate adiabatic attractor $\tau_{s}(x)$, where the dashed part of the curve indicates the zone where the discrepancies between the approximate and exact attractors is more significant.}\label{fig:tau} 
\end{figure}

We consider next the time evolution of the mean values close to the attractor. In order to show this evolution explicitly, we look for a solution of the set of independent equations (\ref{eq_dyn1},\ref{eq_dyn2}) where the variables $x_i^\pm$ are split into their slow (within the attrator) and fast (out of the attractor) parts as
\begin{equation}\label{eq:split1}
 x_i^\pm(t)=x_{i,s}^\pm(t)+ x_{i,f}^\pm(t), \quad i\ge 0.
\end{equation}
In order to ensure that the exact relation between $x$ and $x_i^\pm$ holds, namely Eqs.~(\ref{eq:x01}), we also impose 
\begin{equation}
\label{eq:condf}
 \sum_{i=0}^\infty x_{i,f}^+(t)= \sum_{i=0}^\infty x_{i,f}^-(t)=0.
\end{equation}
Using the fact that $x_{i,s}^\pm$ are solutions to Eqs.~(\ref{eq_dyn1},\ref{eq_dyn2}), we have
\begin{eqnarray}
 \label{eq:f1}
 && \frac{dx_{i,f}^+}{dt} = -x_{i,f}^++x_{i-1,f}^+\alpha_{i-1}(1-x), \quad i\ge 1, \\
 \label{eq:f2} 
 && \frac{dx_{i,f}^-}{dt} = -x_{i,f}^-+x_{i-1,f}^-\alpha_{i-1}(x), \qquad i\ge 1,
\end{eqnarray}
which form a closed set of (linear) equations, provided the conditions (\ref{eq:condf}) are used and $x$ is taken as a parameter. 

The proposed splitting between slow and fast is justified as far as the characteristic time scale evolution to zero of the system of Eqs.(\ref{eq:f1},\ref{eq:f2}) is much smaller than the typical time evolution of $x(t)$, say $t_s \gg 1$ (in MCS) close to the fixed point. In order to estimate the characteristic time evolution of the fast parts $t_{f}$, we analyze Eq.~(\ref{eq:f1}). A similar conclusion can be reached using Eq.(\ref{eq:f2}). The solution to this equation is of the form 
\begin{equation}
\label{eq:sol_expand}
 x_{i,f}^{+}(t;x) = e^{- t} \sum_{k=0}^{\infty} \xi_{i}^{(k)}(x) \frac{t^k}{k!},
\end{equation}
with $\xi_{i}^{(0)}=x_{i,f}^{+}(0)$. The rest of the coefficients $\xi_{i}^{(k)}(x)$ for $k>0$ can be obtained introducing the proposed solution Eq.(\ref{eq:sol_expand}) in Eq.(\ref{eq:f1}) which leads to the recurrence relations $\xi_{i}^{(k+1)}(x)=\alpha_{i-1}(1-x) \xi_{i-1}^{(k)}(x)$ for $i \geq 2$, $\xi_{1}^{(k+1)} (x)= -\alpha_{0}(1-x) \sum_{i=1}^{\infty} \xi_{i}^{(k)}(x)$ and $\xi_{0}^{(k)}(x) = - \sum_{i=1}^{\infty} \xi_{i}^{(k)}(x)$. In principle, it does not seem possible to find a closed solution of these recurrence relations for the coefficients $\xi_{i}^{(k)}$, and consequently we are not able to obtain the full solution $x_{i,f}^{+}(t;x)$. 

As the analysis of the convergence of $x_{i}^{+}(t) \rightarrow x^{+}_{i,s}(t)$, and the time scale of the fast part $x_{i,f}^{+}(t;x)$, is too detailed, we will use an auxiliary aggregated variable, the mean internal age of the nodes in state $+$, i.e. $\tau^{+}(t)=\sum_{i=0}^{\infty} i x_{i}^{+}(t)$ and a similar definition for $\tau^-(t)$. With the proposed variable splitting Eq.(\ref{eq:split1}) we have $\tau^{+}(t)=\tau_{s}^{+}(x(t))+\tau^{+}_{f}(t)$, with $\tau_{s}^{+}(x) = \sum_{i=0}^{\infty} i x_{i,s}^+(x)$ and $\tau_{f}^{+}(t)=\sum_{i=0}^{\infty} i x_{i,f}^{+}(t)$. We then study the approach $\tau^{+}(t) \rightarrow \tau_{s}^{+}(x(t))$, instead of the full $x_{i}^{+}(t) \rightarrow x^{+}_{i,s}(t)$, and assume that it is a good indicator of the time dependence. $\tau_s^+(x)$ can be readily found from Eq.(\ref{eq:at03}), and the advantage is that a closed expression for $\tau_f^+(t)$ can be found using the following relations, that are obtained from the above recurrences for $\xi_{i}^{(k)}$:
\begin{eqnarray}
\label{tau_rec}
\sum_{i=0}^{\infty} i \xi_{i}^{(k+1)} &=& \sum_{i=0}^{\infty} \Big[ (i+1) \alpha_{i} - \alpha_{0} \Big] \xi_{i}^{(k)},\\
(i+1) \alpha_{i} - \alpha_{0} &=& \left(1-\frac{a}{2} \right)i-\left(1-a\right)(1-{x})\big(p_{i}(i+1)-p_{0}\big).
\end{eqnarray}
In the case that ${x} = 1$ or $p_{i} = p_{0}/(i+1)$ the solution is very simple as $\sum_{i=0}^{\infty} i \xi_{i}^{(k)} = \left(1-\frac{a}{2} \right)^{k} \sum_{i=0}^{\infty} i \xi_{i}^{(0)}$ and thus $\tau^{\pm}_{f}(t) = \tau^{\pm}_{f}(0) e^{-at/2}$. The time scale of the fast part is then directly related to the noise intensity $t_{f}=2/a$, and we expect time scale separation as long as $a$ is big enough, as discussed below. 

In order to check these results, we compare in Figs. \ref{fig:tau:a} and \ref{fig:tau:b} the above predicted time evolution of $x(t)$ and $\tau(t) \equiv \tau^+(t)+\tau^-(t)$  with the results of direct numerical simulations of the process given by  Eqs.(\ref{rates1}-\ref{rates4}) using the aging probability $p_{i}=(2+i)^{-1}$ for different initial conditions $x_{i}^{\pm}(0)$. We also plot in Fig. \ref{fig:tau:c} the approximate adiabatic attractor $\tau_{s}(x)$.  One can clearly see in Figs. \ref{fig:tau:b} and \ref{fig:tau:c} the slow and fast contributions to the solution. When $x(t)$ starts at the fixed point $x(0)=1/2$ there is only fast part, while when $x(0)=0$ there is a fast transient and, as $x(t) \rightarrow 1/2$ approaches the fixed point, trajectories follow the theoretical slow part. 
Finally, we conclude that the adiabatic hypothesis is expected to be accurate in the range of parameter values $a$ where $t_{f} \ll t_{s}$, where $t_{s}$ is the time scale of the slow variable $x(t)$, which strongly depends on the type of aging probability $p_{i}$. For example, as shown in \cite{Oriol}, the case $p_{i}=(2+i)^{-1}$ has a critical point $a_{c} = 0.07556...$, thus we will have time scale separation for noise intensities in a window around the critical value $a \sim a_{c}$, where $t_{s} \rightarrow \infty$. From the normal form of Eq.(\ref{eq:x}) $dx/dt \simeq \varepsilon_{1}'(a_{c})(a-a_{c})(x-1/2) + \varepsilon_{3}(a_{c})(x-1/2)^3$ we can identify the slow time scale around the stable fixed points of the dynamics as $t_{s}^{-1} \equiv \varepsilon_{1}'(a_{c})(a-a_{c})$ for $a>a_{c}$ and $t_{s}^{-1} \equiv 2 \varepsilon_{1}'(a_{c})(a_{c}-a)$ for $a<a_{c}$. For the mentioned aging case it is $\varepsilon_{1}'(a_{c}) \simeq 0.372...$, thus we have that $t_{s} > t_{f}$ as long as $0.045 < a \leq 1$.

Although the adiabatic elimination may not give us a perfect accurate dynamical evolution of the variables in the whole parameter region, it is a very good phenomenological approach. Among other properties, it reproduces correctly all critical exponents and finite-size scaling functions of the average value and fluctuations of the global variable, as we will show in the next section.

\section{Markovian reduction}
\label{nonlinear}
\begin{figure}[h!]
\centering
\includegraphics[width=0.7\textwidth]{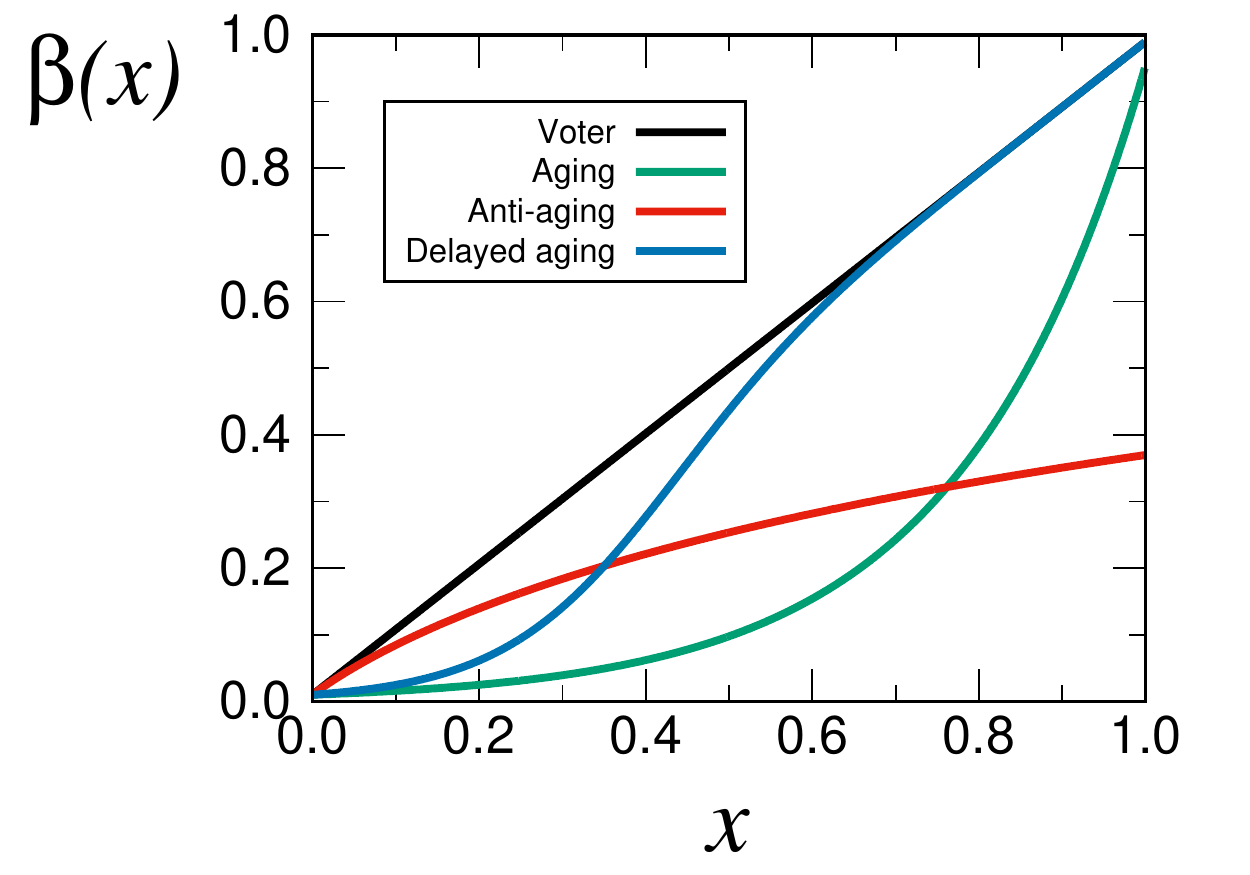}
\caption{Effective non-linear rate $\beta(x)=1/f(1-x)$ for the three different cases considered in Eqs.(\ref{f_aging}, \ref{f_anti}, \ref{f_mixed2}), namely aging, anti-aging and delayed aging with $i_0=5$. In all cases we have set the noise parameter to $a=0.02$. For comparison, we also include the non-aging linear noisy-voter model $p_{i}=1$, $\beta(x)=a/2+(1-a)x$. 
 }\label{fig:rates} 
\end{figure}

In this section we analyze the predictions of the approximated, Markovian, description of the aging voter model based on the master equation (\ref{master-G}) for the global variable $n$ with the effective rates Eqs.(\ref{ratesp},\ref{ratesm}). From a formal point of view, the master equation represents a one-step process \cite{vKampen} in which the variable $n$ can decrease $n\to n-1$ at a rate $\widehat{\Omega}_1$ or increase $n\to n+1$ at a rate $\widehat{\Omega}_2$. Given the form of the effective rates, the process becomes then isomorphic to a voter model without aging in which a randomly selected individual can change its state $-1\to +1$ with rate $\beta(x)=1/f(1-x)$, or change $+1\to -1$ with rate $\beta(1-x)=1/f(x)$, depending in each case on the fraction, $x=n/N$ or $1-x=(N-n)/N$, of individuals in the opposite state. Therefore, the noisy-voter model with linear rates and aging can be approximately replaced by a noisy-voter model without aging and with non-linear rates. The precise form of the individual rate for changing state $\beta(x)=1/f(1-x)$ depends on the activation probability $p_i$. For the non-aging voter model with $p_i=1$ it follows $\beta(x)=\frac{a}{2}+(1-a)x$, whereas the expression for $\beta(x)$ in the aging, anti-aging and delayed aging cases follow readily from the expressions given, respectively, in Eqs.(\ref{f_aging}, \ref{f_anti}, \ref{f_mixed2}). In Fig. \ref{fig:rates} we plot the effective non-linear rates $\beta(x)$ in these different situations. There is a clear qualitative difference between the three cases: while $\beta^{\mathrm{aging}}(x)$ is a convex function with positive second derivative, $\beta^{\mathrm{anti-aging}}(x)$ is concave with negative second derivative and $\beta^{\mathrm{delayed\,aging}}(x)$ has an inflection point, such that is convex for small $x$ and becomes concave for large $x$.

The voter model with non-linear rates has been extensively studied in the literature and it has appeared in several contexts, see \cite{Peralta} and references therein. Typical rates used are of the form $\beta(x)=a+h x^\sigma$ with a non-linearity parameter $\sigma$, and $h$ being an additional parameter. A possible interpretation, suitable when $\sigma=q$ is an integer number is that an agent changes state by imitation if $q$ of its neighbors selected at random hold the opposite state. This so-called $q$-voter model has been the subject of intense research \cite{Castellano2009}. In other applications, $\sigma$ is considered as an adjustable parameter to fit some data and some evidence has been given in problems of language competition that $\sigma>1$ \cite{abrams2003linguistics} , whereas value of $\sigma<1$ corresponding to a probability of imitation above random or a situation of preference for change, have been considered in social impact theory \cite{Nowak1990}. Whatever the interpretation, for $\sigma > 1$ individuals are more reluctant to follow the opinion of the neighbors holding the opposite state, while the opposite is true for $\sigma<1$. The case $\sigma>1$ is somehow reminiscent of the interpretation given to aging as a factor that decreases the likelihood of switching state. The main conclusion of the non-linear voter model with $\sigma>1$ is that partial consensus appears for noise values smaller than a critical value $a<a_c=2^{-\sigma}(\sigma-1)$ through a second-order phase transition. In the non-linearity is very strong, e.g. for $\sigma>5$, the transition between consensus and coexistence becomes first-order with the appearance of a tri-critical point. Due to the aforementioned mapping between the aging and non-linear models it is natural to ask if such a phenomenology also appears for the different aging versions. This is discussed in the next subsection.

\subsection{Phase diagram}
\label{sub:phasediagram}
The master equation (\ref{master-G}) leads to a recurrence relation for the steady-state solution \cite{vKampen}
\begin{equation}
G^{\rm{st}}(n)=G^{\rm{st}}(0)\prod_{i=0}^{n-1}\frac{\widehat{\Omega}_2(i)}{\widehat{\Omega}_1(i+1)}.
\end{equation}
Using the rates given by Eqs.(\ref{ratesp},\ref{ratesm}), it is easy to prove that in the thermodynamic limit $N\to\infty$ the steady-state probability can be written in the large-deviation form~\cite{touchettelarge2009} $G^{\rm{st}}(n)\sim e^{-NV(n/N)}$, where the potential function $V(x)$ is given by
\begin{equation}
\label{potentialfromf}
V(x)=\int^x dx \log\left[\frac{x f(1-x)}{(1-x)f(x)}\right].
\end{equation}
Note that, as the extrema of $V(x)$ coincide with the exact stationary states $x_\mathrm{st}$ obtained from Eq.(\ref{eq:xest}), the adiabatic approximation does not modify them or introduce new steady states. However, the advantage of considering this potential function is that, in the thermodynamic limit, the stable phases are associated to absolute minima of the potential $V(x)$.  Alternatively, the stable phases can be considered as fixed points of the dynamical equation $\frac{dx}{dt}=-\frac{dV(x)}{dx}$.

For the aforementioned cases of the activation probability, Eqs.(\ref{f_aging},\ref{f_anti}), we obtain the following potential functions:
\begin{eqnarray}
V^{\rm{aging}}(x)&=&\Phi_1(x)+\Phi_1(1-x), \\ \Phi_1(x)&=&x\left[\log(x)-\frac{\kappa}{2}\log\left(\frac{2}{a}\right)x\right],\nonumber\\
V^{\rm{anti-aging}}(x)&=&\Phi_2(x)+\Phi_2(1-x), \\
\Phi_2(x)&=&x\log(x)+\kappa^{-1}\left(\log\left[1-g(x)\right]\log\left[g(x)\right]+{\rm Li}_2\left[g(x)\right]\right),\nonumber
\end{eqnarray}
where ${\rm Li}_2(z)$ is the polylogarithm function. For the delayed-aging case, we have not been able to obtain an analytically closed formula for $V^{\rm{delayed\,aging}}(x)$ and we need to perform numerically the integral Eq.(\ref{potentialfromf}).

\begin{figure}[h!]
\centering
\includegraphics[width=0.7\textwidth]{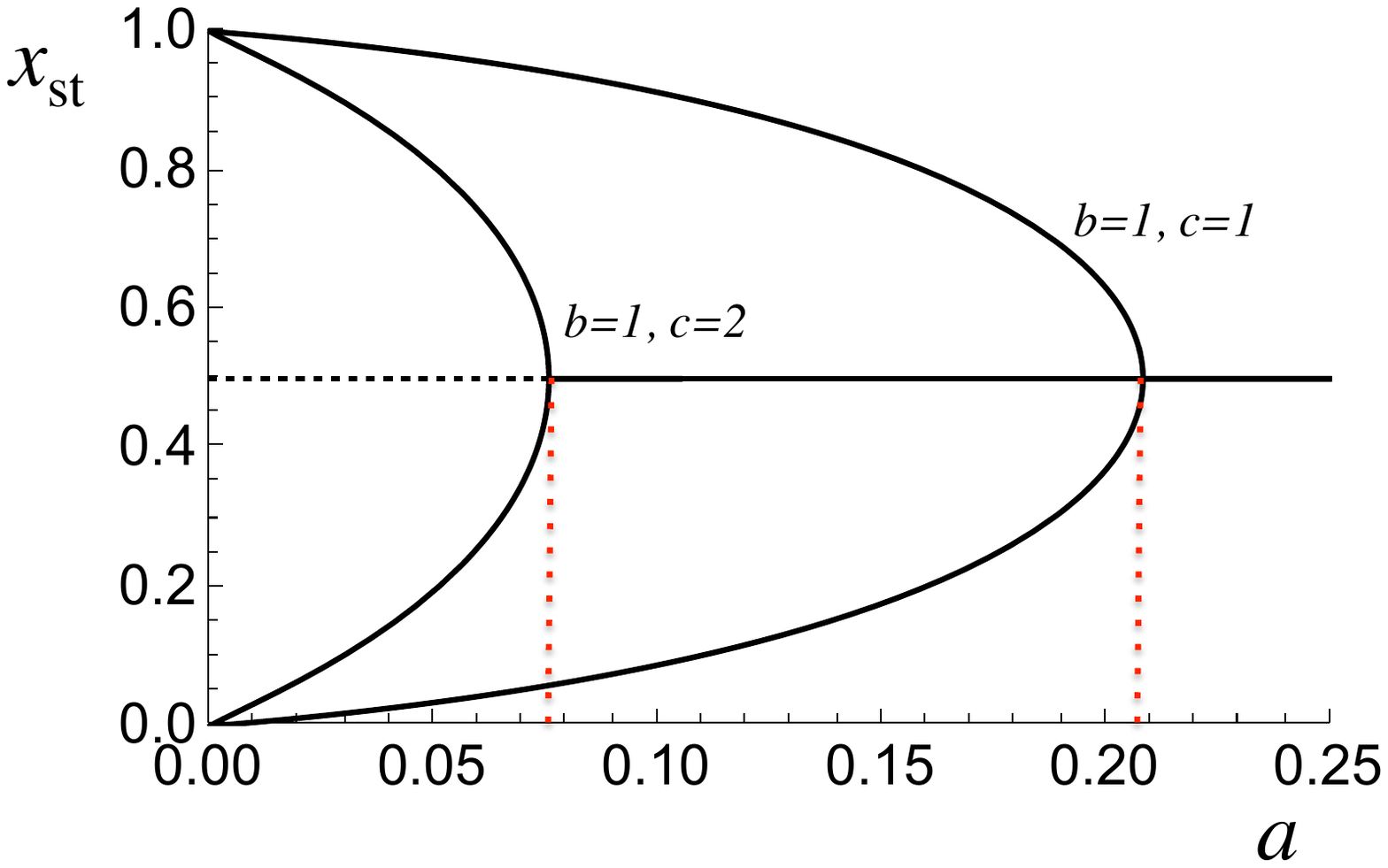}
\label{fig:phase} 
\caption{Stable steady states $x_\mathrm{st}$ of the fraction of agents holding the $+1$ state in the case of aging probabilities given by Eq.(\ref{eq:paging}), coming from the absolute minima of the potential Eq.(\ref{potentialfromf}). The solid curve in the right corresponds to $b=c=1$, as indicated, where $f^{\mathrm{aging}}(x)$ is given in Eq.(\ref{f_aging}), and the solid curve in the left to $b=1,\,c=2$ for which one can obtain $f^{\mathrm{aging}}(x) = \frac{\left( 2/a \right)^{\frac{1-(1-a)x}{1-a/2}}-2}{a+2(1-a)x}$. The dashed black line corresponds to the unstable symmetric solution $x=1/2$. Note the continuous phase transition at $a=a_c$.}
\end{figure}

As, in all cases, the potential $V(x)$ is symmetric around $x=1/2$, it seems convenient to write it in terms of the {\slshape magnetization} $m \equiv 2 x - 1\in[-1,1]$ as a symmetric order parameter. The stability of the $m=0$ state is determined by the expansion of the potential 
\begin{equation}
\label{Vm}
V(m) = V_2m^2 + V_4m^4 + \dots, \quad V_k=\frac{1}{2^k k!}\left.\frac{d^kV(x)}{dx^k}\right\vert_{x=1/2}.
\end{equation} Depending on the sign of the coefficients, one can determine if the extremum is stable or unstable, if there is a critical point where stability changes, and the number of new extrema that appear at the transition. It is easy to show that the coefficients can be determined as a function of $f(x)$ as follows
\begin{equation}
\label{coeff1}
V_n= \frac{1}{n!2^{n-1}} \left(2^{n-1}(n-2)! + d_{n} \right),
\end{equation}
where $d_{n}=(-1)^{n-1}\left.\frac{d^{n-1}\log f(x)}{dx^{n-1}} \right\vert_{x=1/2}$. For the first two coefficients we have $V_2=\frac{1}{4}\left(2+ d_2 \right)$ and $V_4 = \frac{1}{192}\left(16+ d_4 \right)$. Moreover, for the particular case of aging given by Eq.(\ref{f_aging}), we find $V_{2} =\frac12\left[1-\frac{1-a}{2-a} \log \left(\frac{2}{a} \right)\right]$, and $d_{n > 2}=0$. As the coefficient $V_2$ changes sign at $a=a_c$, defined as $\frac{1-a_c}{2-a_c} \log \left(\frac{2}{a_c} \right)=1$, or $a_{c} = 0.2081\dots$, we conclude\footnote{In Ref.~\cite{Oriol} we used $b=1,\,c=2$ and obtained a critical value $a_c=0.07556\dots$} that the extremum $m=0$ is stable if $a > a_{c}$ and unstable if $a < a_{c}$. Since $V_4=1/12>0$, two new stable solutions appear for $a < a_{c}$ at a pitchfork bifurcation. For the anti-aging case Eq.(\ref{f_anti}), it is $V_{2}=\frac{1}{2}\left[1+(\log 2-1)(1-a)(2-a)\right] > 0$ for all values of $a\in[0,1]$, thus the symmetric solution $m=0$ is always stable~\cite{Ozaita}. In Fig. \ref{fig:phase} we plot the phase diagram for the aging case. 

\begin{figure}[h!]
\centering
\subfloat[]{\label{fig:fixed:a}\includegraphics[width=0.45\textwidth]{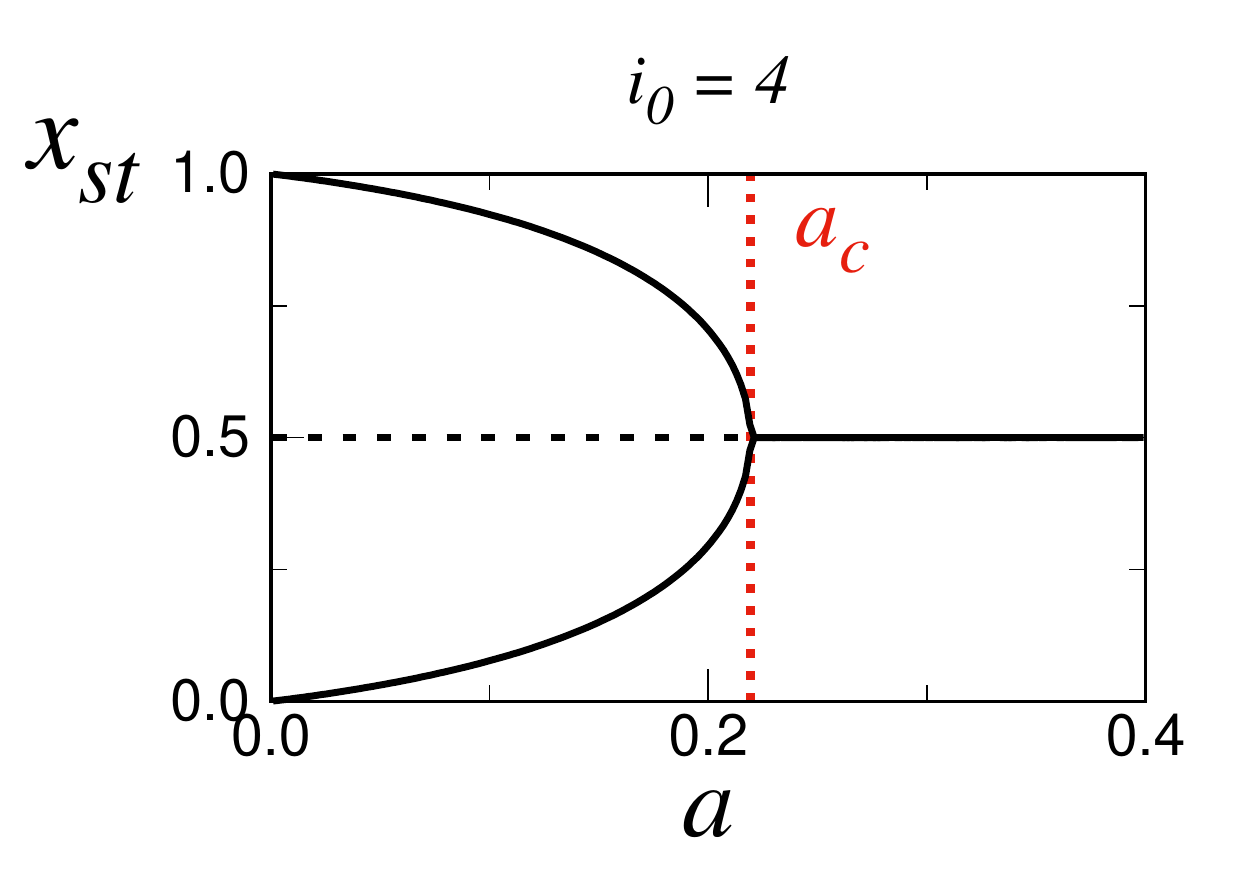}}
\subfloat[]{\label{fig:fixed:b}\includegraphics[width=0.45\textwidth]{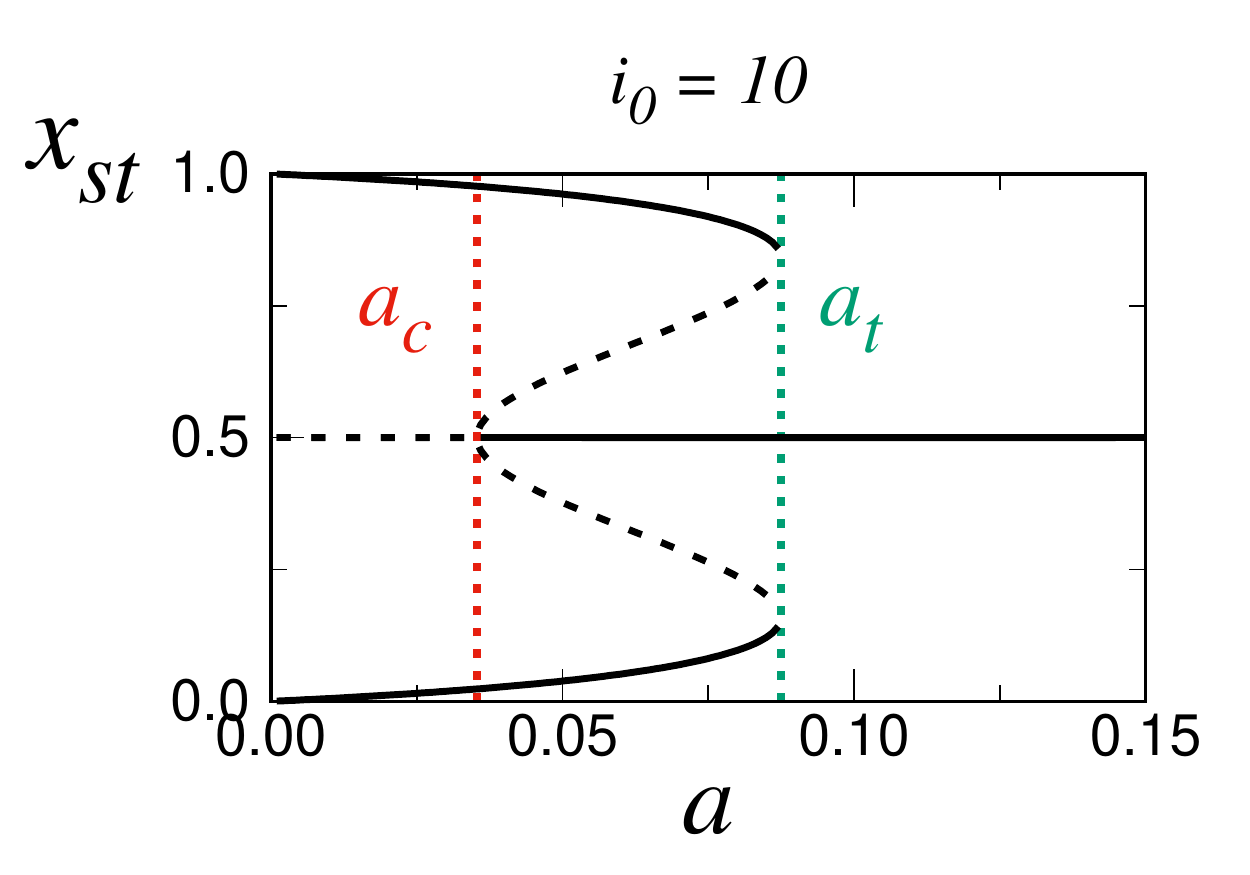}}

\subfloat[]{\label{fig:fixed:c}\includegraphics[width=0.45\textwidth]{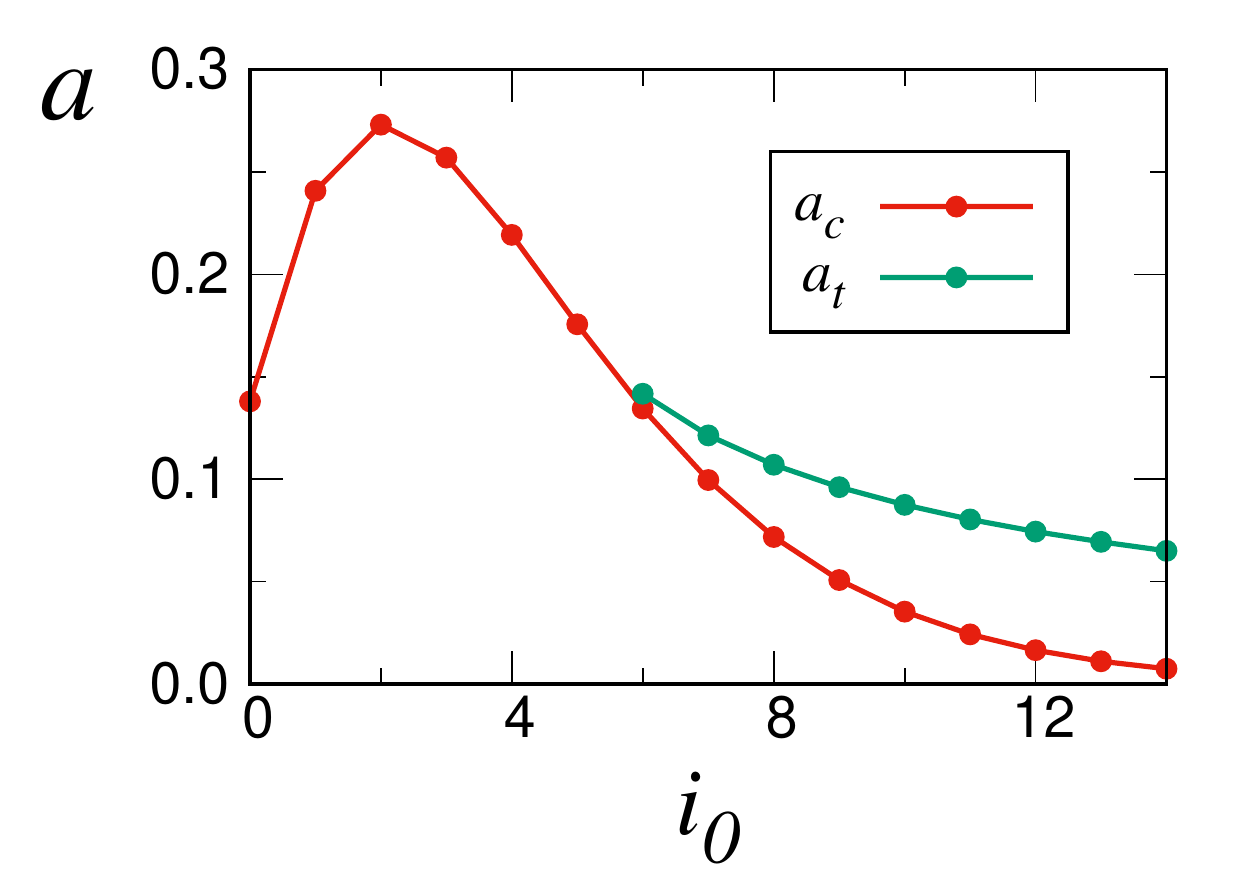}}
\caption{Steady states of the global variable $x$ for the voter model with delayed aging, with $p_{i\leq i_{0}}=1$ and $p_{i>i_{0}}=1/(i-i_{0})$ as a function of $a$ for different values of $i_{0}$ indicated in the figure. The panel (c) shows $a_{c}$ (red) and $a_{t}$ (green) as a function of $i_{0}$. }\label{fig:fixed} 
\end{figure}

The delayed aging case is more complicated and we will not give explicit details of the analysis, as the resulting expressions are too long. What we obtain is that the coefficient $V_2$ also changes sign at a critical point $a_c$ for all values of $i_0$, as in the aging case. The difference is that the sign of the coefficient $V_4$ at the critical point $a_{c}$ depends on the value of $i_{0}$. For $i_{0} < i_{0}^{c}$, we have $V_4>0$ (supercritical pitchfork bifurcation), while for $i_{0} > i_{0}^{c}$, it is $V_4<0$ (subcritical pitchfork bifurcation), see Fig. \ref{fig:fixed}. The supercritical case is equivalent to the aging case, while for the subcritical one we have two critical points, the pitchfork one $a_{c}$ where $V_2$ changes sign, and another one $a_{t} > a_{c}$ of a saddle node transition. This gives three zones: (i) for $a<a_{c}$ there are three fixed points, two stable and the symmetric unstable, (ii) for $a_{c}<a<a_{t}$ there are five fixed points, the symmetric becomes stable and two additional unstable fixed points add to the case (i), (iii) for $a>a_{t}$ the pairs of stable-unstable fixed points disappear in a saddle node transition, leaving only one fixed point, the symmetric which remains stable, see Fig. \ref{fig:fixed}. This phenomenology is similar to the one observed in the non-linear voter model for $\sigma>5$~\cite{Peralta}.

\subsection{Fluctuations around fixed points}

\begin{figure}[h!]
\centering
\subfloat[]{\label{fig:fluct:a}\includegraphics[width=0.45\textwidth]{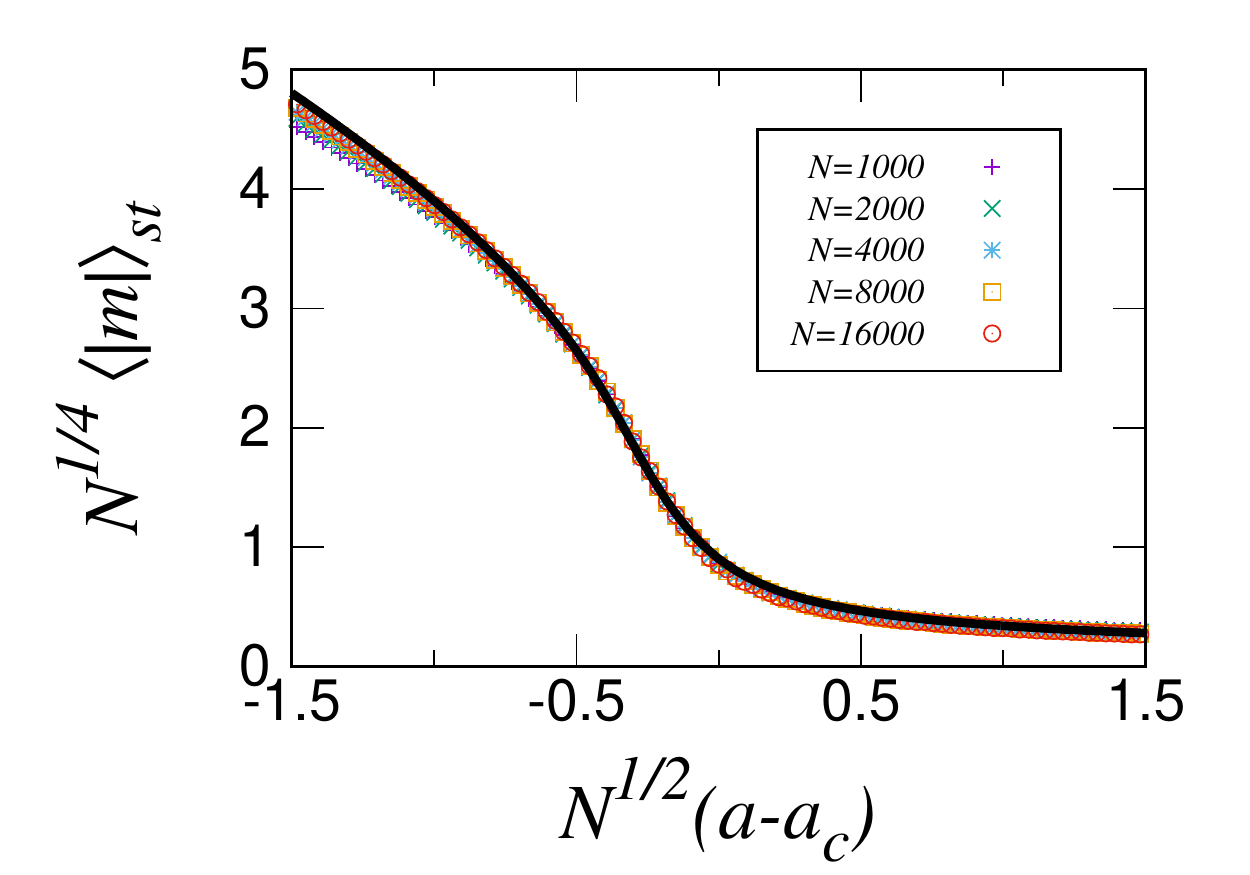}}
\subfloat[]{\label{fig:fluct:b}\includegraphics[width=0.45\textwidth]{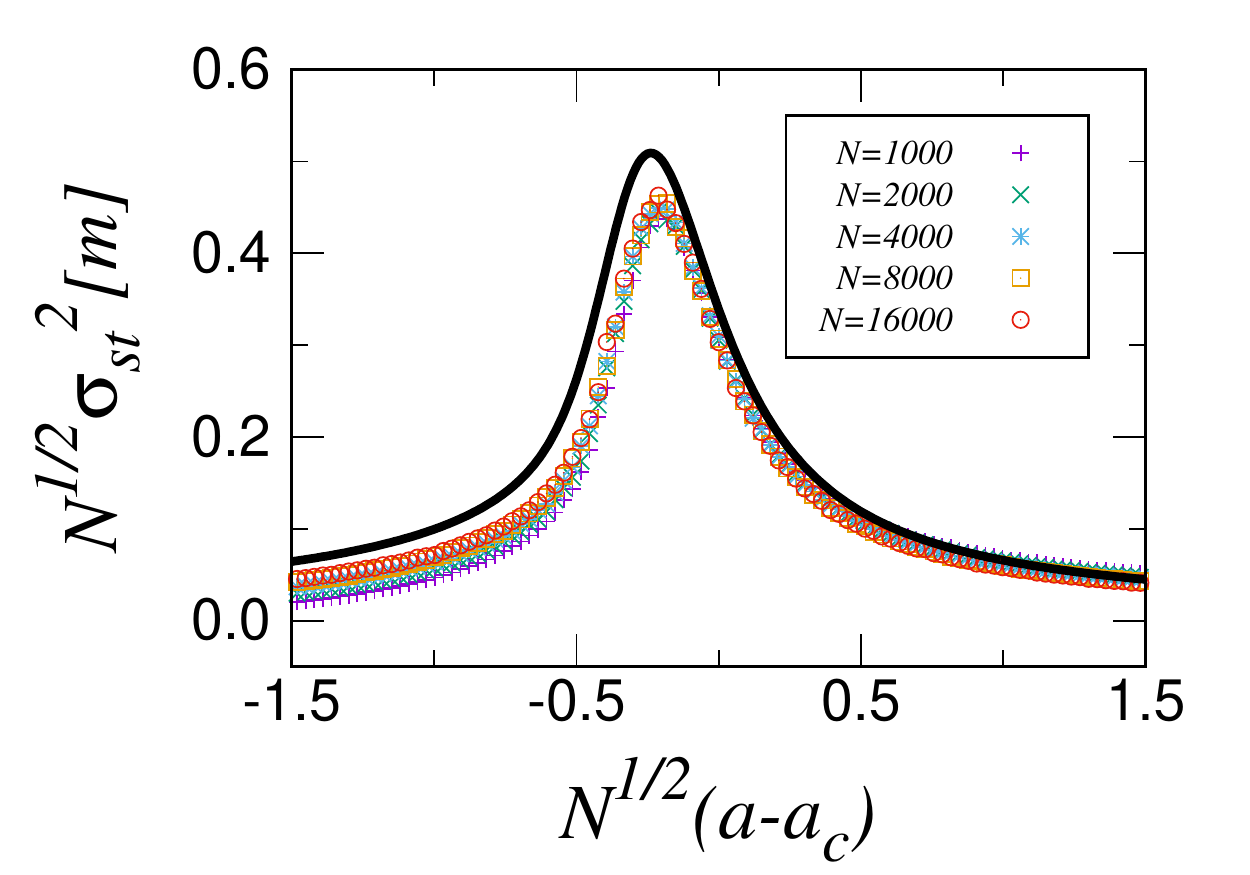}}

\subfloat[]{\label{fig:fluct:c}\includegraphics[width=0.45\textwidth]{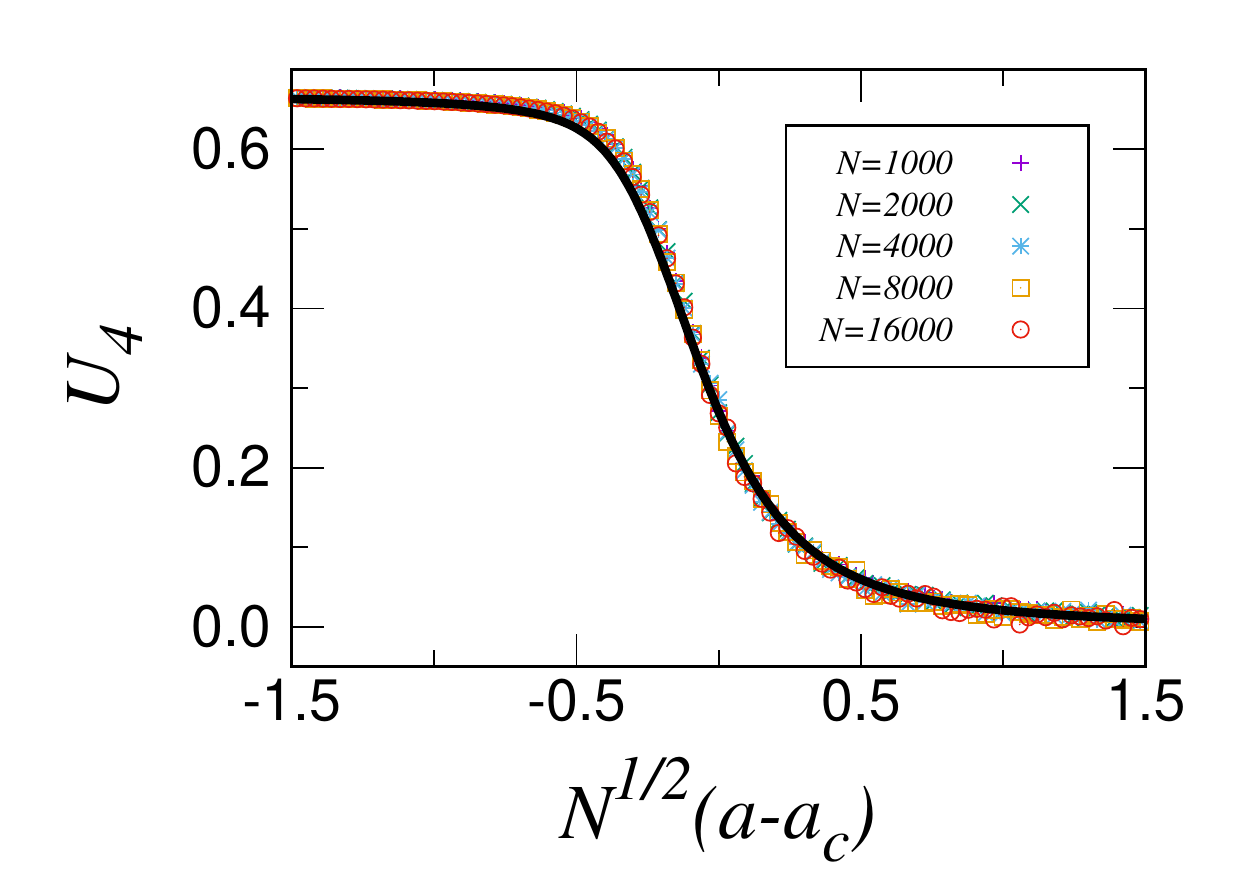}}
\caption{Magnetization $\langle \vert m \vert \rangle$, variance $\sigma^{2}[m] = \langle m^2\rangle - \langle \vert m \vert \rangle^2$, and Binder cumulant $U_4 = 1-\langle m^4\rangle/(3\langle m^2\rangle^2)$ as a function of noise $a-a_{c}$, for the aging model with an activation probability $p_{i}=1/(2+i)$, for which one can obtain the coefficients $c=2.643$ and $c_{4}=0.0852$ of the potential Eq.(\ref{potential_coeff}). The results of numerical simulation are plotted as points with different colors, corresponding to different system sizes $N$, averaged over $10^6$ Monte Carlo steps, while the corresponding theoretical scaling function obtained from Eq.(\ref{scaling_funct}) is the solid black line.}\label{fig:fluct} 
\end{figure}

Given the form $G^{\rm{st}}(n)\sim e^{-NV(n/N)}$, the dependence with system size $N$ of the moments of the magnetization $\left\langle m^{k} \right\rangle_{\mathrm{st}}$ near the critical point $a=a_c$ can be obtained~\cite{Peralta} from the expansion of the coefficients of $V(m)$:
\begin{eqnarray}
\label{potential_coeff}
V_{2}(a) \simeq c (a-a_{c}), \quad V_{4}(a) \simeq c_{4},
\end{eqnarray}
as $\left\langle m^{k} \right\rangle_{\mathrm{st}}=N^{-k/4} \phi_{k} \left[ N^{1/2} (a-a_{c}) \right]$ with a scaling function $\phi_{k}[s]$:
\begin{equation}
\label{scaling_funct}
\phi_{k}[s]=\frac{\int_{-\infty}^{\infty} z^{k} e^{-c s z^2-c_{4} z^4} dz}{\int_{-\infty}^{\infty} e^{-c s z^2-c_{4} z^4} dz}.
\end{equation}
In Fig. \ref{fig:fluct} we compare the scaling behavior predicted by this theoretical analysis for the absolute magnetization $\langle |m|\rangle$, the Binder cumulant $U_4 = 1-\langle m^4\rangle/(3\langle m^2\rangle^2)$ and the variance $\sigma^{2}[m]=\langle m^2\rangle-\langle |m|\rangle^2$ with the results coming from a numerical simulation of the aging model at different system sizes, using an activation probability $p_{i}=1/(2+i)$. It can be seen that in all cases the theoretical scaling curves match remarkably well those of the simulations, validating the Markovian reduction introduced in this paper. In the thermodynamic limit, these scaling laws lead to $\langle \vert m \vert \rangle \sim (a_{c}-a )^{1/2}$ and $\chi \equiv N \sigma^{2}[m] \sim \vert a_{c}-a \vert ^{-1}$ for the susceptibility, which correspond to the mean-field critical exponents $\beta=1/2$, $\gamma=1$.

\section{Conclusions}\label{sec_conc}

In this work we have studied a non-Markovian binary-state model, which is constructed by including memory effects (aging) in the noisy-voter model. By means of a theoretical analysis, we have proved the the non-Markovian model, whose individual rates are linear with respect to the density of neighbor agents holding the opposite state, can be reduced to a non-linear noisy-voter model which is Markovian.

This Markovian reduction is always valid, as long as the noise is nonzero $a \neq 0$, for determining the deterministic steady state values. For this reason, we were able to show that all the phenomenology found in the critical behavior of the non-linear noisy-voter model is also observed for the non-Markovian linear model, which includes induced continuous phase transitions, discontinuous transitions, tricritical behavior, etc. 

With respect to the dynamics and fluctuations of the model we have also shown that, in most of the cases, the same Markovian reduction is possible. This happens when the evolution of the system towards the steady states is controlled by the global fraction of agents with a given opinion $x$ (regardless the age). In the sense that, the evolution of the number of agents with given opinion and age (fast variables) rapidly slaves to the value of $x$ (slow variable). This adiabatic elimination of the age variables is valid as long as there is time scale separation, which strongly depends on the type of aging and on the noise intensity $a$. The adiabatic elimination technique allowed us to determine the finite-size scaling functions of the fluctuations and the dynamics of the model, which are impossible to obtain in the full non-Markovian description.

In the absence of noise $a=0$, the general picture depicted above does not apply, in general. At an absorbing or consensus state of the voter model with aging, for instance, the number of individuals in a given state (regardless the age) does not change while the number of agents with the consensus opinion and with a given age changes continuously due to aging. Hence, the splitting in terms of fast and slow variables is not possible and, we have to resort to other analytical techniques, as studied in ~\cite{Peralta2019}.

\section*{Acknowledgements}
Partial financial support has been received from the Agencia Estatal de Investigaci\'on (AEI, MCIU, Spain) and Fondo Europeo de Desarrollo Regional (FEDER, UE), under Project PACSS (RTI2018-093732-B-C21/C22) and the Maria de Maeztu Program for units of Excellence in R\&D (MDM-2017-0711). A.F.P. acknowledges support by the Formaci\'on de Profesorado Universitario (FPU14/00554) program of Ministerio de Educaci\'on, Cultura y Deportes (MECD) (Spain). 

\appendix
\section{The determination of the exact attractor}\label{app:attractor}

In this Appendix we obtain the exact dynamical attractor $x_{i,s}^{\pm}(x(t))$ and we derive the time evolution of the global variable $x(t)$, comparing altogether with an adiabatic elimination $dx/dt \approx 0$.

For this purpose, we try a solution $x_{i,s}^{\pm}(x)$ of Eqs.(\ref{atractor1},\ref{atractor2}) in power series around a fixed point $\widetilde{x}$, $\left. \frac{dx}{dt} \right|_{x=\widetilde{x}}=0$:
\begin{equation}
\label{attractor_expand}
x_{i,s}^{\pm}(x) = \sum_{n=0}^{\infty} g_{i}^{\pm(n)} (x-\widetilde{x})^{n},
\end{equation}
where $g_{i}^{\pm(n)}$ are the coefficients of the expansion. Note that, due to the symmetries of Eqs.(\ref{atractor1},\ref{atractor2}), we have that $x_{i,s}^{-}(x)=x_{i,s}^{+}(1-x)$, thus the coefficients $g_{i}^{-(n)}$ can be obtained from $g_{i}^{+(n)}$, and vice versa. As the solution has to fulfill the relations Eq.(\ref{eq:x01}), the coefficients are restricted as $\sum_{i=0}^{\infty} g_{i}^{+(0)} = \widetilde{x}$, $\sum_{i=0}^{\infty} g_{i}^{-(0)} = 1-\widetilde{x}$, $\sum_{i=0}^{\infty} g_{i}^{+(1)} = 1$, $\sum_{i=0}^{\infty} g_{i}^{-(1)} = -1$ and $\sum_{i=0}^{\infty} g_{i}^{\pm(n)} = 0$ for $n \geq 2$. After introducing the expansion Eq.(\ref{attractor_expand}) in Eq.(\ref{atractor3}) we obtain the expanded equation for the global $x$:
\begin{equation}
\label{eqx_expand}
\frac{dx}{dt} = \sum_{n=1}^{\infty} \varepsilon_{n} (x-\widetilde{x})^{n},
\end{equation}
where $\varepsilon_{n}$ depend on $g_{i}^{\pm(n)}$. 

For the symmetric fixed point $\widetilde{x}=1/2$, the coefficients fulfill the symmetry relations $\varepsilon_{n}=0$ and $g_{i}^{+(n)}=-g_{i}^{-(n)}$ for $n$ odd, and $g_{i}^{+(n)}=g_{i}^{-(n)}$ for $n$ even, obtained easily from the restriction $x_{i,s}^{-}(x)=x_{i,s}^{+}(1-x)$. Introducing the expansion Eqs.(\ref{attractor_expand},\ref{eqx_expand}) in Eq.(\ref{atractor1}) and equating the different orders of $O(x-\widetilde{x})$, we find the following recurrence relations for the coefficients $g_{i}^{+(n)}$ of the first $n=0,1,2,3$:
\begin{eqnarray}
\label{rec1}
0 &=& - g_{i}^{(0)} + g_{i-1}^{(0)} a_{i-1}, \\
\label{rec2}
g_{i}^{(1)} \varepsilon_1 &=& - g_{i}^{(1)} + g_{i-1}^{(1)} a_{i-1} - g_{i-1}^{(0)} a'_{i-1}, \\
\label{rec3}
2 g_{i}^{(2)} \varepsilon_1 &=& - g_{i}^{(2)} + g_{i-1}^{(2)} a_{i-1} - g_{i-1}^{(1)} a'_{i-1}, \\
\label{rec4}
g_{i}^{(1)} \varepsilon_{3} + 3 g_{i}^{(3)} \varepsilon_1 &=& - g_{i}^{(3)} + g_{i-1}^{(3)} a_{i-1} - g_{i-1}^{(2)} a'_{i-1},
\end{eqnarray}
where $a_{i} \equiv \alpha_{i}(1/2)$, $a_{i}' \equiv \left. \frac{d}{dx}\alpha_{i}(x) \right|_{x=1/2}$ and we have simplified notation by removing the superscript $+$, such that $g_i^{(n)}$  stands for $g_i^{+(n)}$. The first recurrence relation Eq.(\ref{rec1}) leads trivially to the stationary solution Eqs.(\ref{eq:xi1},\ref{eq:x03}) at the symmetric fixed point $\widetilde{x}=1/2$, this is:
\begin{equation}
\label{rec1_sol}
g_{i}^{(0)} = g_{0}^{(0)} \prod_{k=0}^{i-1} a_{k}, \hspace{1.0cm} g_{0}^{(0)} = \frac{1}{2 f},
\end{equation} 
with $f=1+\sum_{i=1}^{\infty}\prod_{k=0}^{i-1} a_{k}$. The second recurrence relation Eq.(\ref{rec2}) leads to:
\begin{equation}
\label{rec2_sol1}
g_{i}^{(1)} = \frac{g_{0}^{(1)}}{(1+\varepsilon_{1})^i} \prod_{k=0}^{i-1} a_{k} - g_{0}^{(0)} \prod_{k=0}^{i-1} a_{k} \sum_{k=0}^{i-1} \frac{a_{k}'}{a_{k}} \frac{1}{(1+\varepsilon_{1})^{i-k}},
\end{equation} 
with
\begin{equation}
\label{rec2_sol2}
g_{0}^{(1)} = \frac{1+g_{0}^{(0)}\sum_{i=1}^{\infty}\prod_{k=0}^{i-1} a_{k} \sum_{k=0}^{i-1} \frac{a_{k}'}{a_{k}} \frac{1}{(1+\varepsilon_{1})^{i-k}}}{1
+\sum_{i=1}^{\infty} \frac{1}{(1+\varepsilon_{1})^{i}} \prod_{k=0}^{i-1} a_{k}}.
\end{equation} 

If we introduce the expansion Eq.(\ref{attractor_expand}) in Eq.(\ref{eq:x01}) we get the relations between $\varepsilon_{1,3}$ and the first four coefficients $g_{i}^{(0,1,2,3)}$ for the symmetric fixed point $\widetilde{x}=1/2$:
\begin{eqnarray}
\label{epsilon1}
\varepsilon_{1} &=& 2 \sum_{i=0}^{\infty} \left[g_{i}^{(0)} b'_{i} - g_{i}^{(1)} b_{i} \right], \\
\label{epsilon3}
\varepsilon_{3} &=& 2 \sum_{i=0}^{\infty} \left[g_{i}^{(2)} b'_{i} - g_{i}^{(3)} b_{i} \right],
\end{eqnarray}
with $b_{i} \equiv \beta_{i}(1/2)$ and $b'_{i} \equiv \left. \frac{d}{dx}\beta_{i}(x) \right|_{x=1/2}$. Eq.(\ref{epsilon1}) is an implicit equation for $\varepsilon_{1} = L(a,\varepsilon_{1})$ as a function of the noise intensity $a$. In order to analyze the dependence $\varepsilon_{1}(a)$, we will assume that there is a critical point $a = a_{c}$ where the result of the equation is $\varepsilon_{1}=0$, i.e. $L(a_{c},0)=0$. We can then expand $L(a,\varepsilon_{1})$ around this point:
\begin{equation}
\label{epsilon_crit}
\varepsilon_{1} \approx \frac{\partial L}{\partial a} (a-a_{c}) + \frac{\partial L}{\partial \varepsilon_{1}} \varepsilon_{1},
\end{equation}
where the derivatives are evaluated at $a=a_{c}$ and $\varepsilon_{1}=0$. The solution close to $a \approx a_{c}$ can be obtained as $\varepsilon_{1} =\frac{\partial_{a} L}{1-\partial_{\varepsilon_{1}}L} (a-a_{c})$. 

In the bare adiabatic elimination Eq.(\ref{eq:or0}), however, the coefficients $g_{i}^{(n)}$ of the attractor are determined by imposing the r.h.s. of Eqs.(\ref{rec1}-\ref{rec4}) equal to zero, which is accurate only when $\varepsilon_{n} \approx 0$, i.e. very slow $x(t)$ (exact for $g_{i}^{(1)}$, $g_{i}^{(2)}$ at $\varepsilon_{1} = 0$, $a=a_{c}$). The equation for $\varepsilon_{1}$ becomes explicit $\varepsilon_{1} = L(a,0)$ with the same function $L$ of the exact attractor. Near $a \approx a_{c}$ we have $\varepsilon_{1} \approx \frac{\partial L}{\partial a} (a-a_{c})$ and the adiabatic elimination neglects $\frac{\partial L}{\partial \varepsilon_{1}} \approx 0$. The adiabatic elimination may not be perfectly accurate but we note that the two procedures, i.e. obtaining the exact and adiabatic attractors, give the same normal form of Eq.(\ref{eqx_expand}), this is:
\begin{equation}
\label{normal_form}
\frac{dx}{dt} \simeq \varepsilon_{1}'(a_{c}) (a-a_{c}) (x-\widetilde{x}) + \varepsilon_{3}(a_{c}) (x-\widetilde{x})^3 + O(x-\widetilde{x})^5, \
\end{equation}
where we write the explicit dependence of the coefficients $\varepsilon_{n}(a)$ with respect to the noise $a$, and $\varepsilon_{1}'(a_{c}) = \left. \frac{d}{da} \varepsilon_{1} \right|_{a=a_{c}}$. The two methodologies give the same critical exponents, scaling functions, symmetries and phenomenology, although the numerical value of the coefficients $\varepsilon_{1}'(a_{c})$ and $\varepsilon_{3}(a_{c})$ may not be correct for the adiabatic elimination method. It is also important to realize that, although the methodology presented here relies on an expansion around a critical point $a \approx a_{c}$, its validity extends over a wide range of parameter values. This is, the normal form Eq.(\ref{normal_form}) describes accurately the time evolution of $x(t)$ for almost all $a \in (0,1)$.

We will now compute the difference between the exact and the approximate adiabatic coefficients of the normal form, i.e. $\varepsilon_{1}'(a_{c})$, $\varepsilon_{3}(a_{c})$. For this reason we redefine the expansion as $g_{i}^{(n)} = s_{i}^{(n)} + d_{i}^{(n)}$, where we split the coefficients with the contribution of the adiabatic elimination $s_{i}^{(n)}$ and the correction $d_{i}^{(n)}$. The adiabatic coefficients $s_{i}^{(n)}$ can be obtained solving the recurrence relations Eqs.(\ref{rec1}-\ref{rec4}) with $\varepsilon_{n}=0$, $\forall n$, i.e. $dx/dt \approx 0$, while the corrections $d_{i}^{(n)}$ are just the difference between the exact solution $g_{i}^{(n)}$ and the adiabatic approximation $s_{i}^{(n)}$. For determining $\varepsilon_{1}'(a_{c})$ we have to calculate the first term of $g_{i}^{(1)}=s_{i}^{(1)}+d_{i}^{(1)}$ of Eqs.(\ref{rec2_sol1},\ref{rec2_sol2}) for small $\varepsilon_{1}$ at the critical point $a=a_{c}$ (the contribution with $\varepsilon_{1} = 0$ corresponds the adiabatic approximation $s_{i}^{(1)}$), this is:
\begin{eqnarray}
\label{coefficients1_s1}
d^{(1)}_{i} &=&  d^{(1)}_{0} \prod_{k=0}^{i-1} \alpha_{k} + \frac{\varepsilon_{1}}{2 f} \prod_{k=0}^{i-1} a_{k} \sum_{k=0}^{i-1} \frac{a'_{k}}{a_{k}} (i-k), \\
\label{coefficients2_s1}
d^{(1)}_{0} &=&  -\frac{\varepsilon_{1}}{2 f^2} \sum_{i=1}^{\infty} \prod_{k=0}^{i-1} a_{k} \sum_{k=0}^{i-1} \frac{a'_{k}}{a_{k}} (i-k).
\end{eqnarray}
Introducing this $d_{i}^{(1)} \propto \varepsilon_{1}$ in Eq.(\ref{epsilon1}) gives us the term $\partial_{\varepsilon_{1}} L$ of Eq.(\ref{epsilon_crit}):
\begin{equation}
\label{G_eps1}
\frac{\partial L}{\partial \varepsilon_{1}} =  \frac{1}{f} \sum_{i=1}^{\infty} \left[ \frac{1}{f} - b_{i} \right] \prod_{k=0}^{i-1} a_{k} \sum_{k=0}^{i-1} \frac{a'_{k}}{a_{k}} (i-k).
\end{equation}

In order to calculate $\varepsilon_{3}(a_{c})$ at the critical point $a=a_{c}$, we impose $\varepsilon_{1}(a_{c})=0$ in Eqs.(\ref{rec2}-\ref{rec4}) and we realize that the coefficients $g_{i}^{(1)} = s_{i}^{(1)}$, $g_{i}^{(2)} = s_{i}^{(2)}$ of the adiabatic elimination are exact ($d_{i}^{(1)}=0$, $d_{i}^{(2)}=0$). Thus, we only need to find $g_{i}^{(3)} = s_{i}^{(3)} + d_{i}^{(3)}$ from the recurrence relation Eq.(\ref{rec4}) and introduce it in Eq.(\ref{epsilon3}). This leads to the solution:
\begin{eqnarray}
\label{coefficients1_s3}
d^{(3)}_{i} &=&  d^{(3)}_{0} \prod_{k=0}^{i-1} a_{k} + \frac{\varepsilon_{3}}{2 f} \prod_{k=0}^{i-1} a_{k} \sum_{k=0}^{i-1} \frac{a'_{k}}{a_{k}} (i-k), \\
\label{coefficients2_s3}
d^{(3)}_{0} &=&  -\frac{\varepsilon_{3}}{2 f^2} \sum_{i=1}^{\infty} \prod_{k=0}^{i-1} a_{k} \sum_{k=0}^{i-1} \frac{a'_{k}}{a_{k}} (i-k),
\end{eqnarray}
which has the same form as Eqs.(\ref{coefficients1_s1}, \ref{coefficients2_s1}) with $d_{i}^{(3)} \propto \varepsilon_{3}$. Consequently the exact equation for $\varepsilon_{3}$ Eq.(\ref{epsilon3}) deviates from the adiabatic approximation $d_{i}^{(3)}=0$ in an equivalent way as Eq.(\ref{epsilon_crit}), i.e. $\varepsilon_{3} =\varepsilon_{3}(\mathrm{adiabatic}) + \frac{\partial L}{\partial \varepsilon_{1}} \varepsilon_{3} \rightarrow \varepsilon_{3}= \frac{\varepsilon_{3}(\mathrm{adiabatic})}{1-\partial_{\varepsilon_{1}}L}$. 

Finally, we conclude that the normal form of the exact attractor Eq.(\ref{normal_form}) is the same as the one of the crude adiabatic elimination after multiplying the whole equation by the pre-factor ${\cal C} = (1-\partial_{\varepsilon_{1}} L)^{-1}$. Where, the partial derivative $\partial_{\varepsilon_{1}} L$ has to be evaluated using Eq.(\ref{G_eps1}) at the critical point $x=1/2,\,a=a_{c}$ with $\varepsilon_{1}=0$. For example in the case of Fig. \ref{fig:tau}, i.e. with aging $p_{i}=(2+i)^{-1}$ and $a_{c}\simeq 0.07556$, we obtain a pre-factor ${\cal C} = 0.316...$ In section \ref{nonlinear}, see Figure \ref{fig:fluct}, we test the accuracy of the finite-size scaling functions calculated with the coefficients extracted using the crude adiabatic elimination, which leads to a remarkable good agreement compared to numerical simulation. This is because this pre-factor is not important for calculating the scaling functions, but it is indeed important for the dynamical evolution of the slow variable $x(t)$, as shown in Fig. \ref{fig:tau}.

\section*{References}

\begin{thebibliography}{10}
\expandafter\ifx\csname url\endcsname\relax
  \def\url#1{\texttt{#1}}\fi
\expandafter\ifx\csname urlprefix\endcsname\relax\def\urlprefix{URL }\fi
\expandafter\ifx\csname href\endcsname\relax
  \def\href#1#2{#2} \def\path#1{#1}\fi

\bibitem{voter-election}
J.~Fern\'andez-Gracia, K.~Suchecki, J.~J. Ramasco, M.~{San Miguel}, V.~M.
  Egu\'{\i}luz, Is the voter model a model for voters?, Phys. Rev. Lett. 112
  (2014) 158701.

\bibitem{Kononovicius2018}
A.~Kononovicius, Modeling of the parties' vote share distributions, Acta
  Physica Polonica A 133~(6) (2018) 1450--1458.

\bibitem{Kononovicius2019}
A.~Kononovicius, Compartmental voter model, Journal of Statistical Mechanics:
  Theory and Experiment 2019~(10) (2019) 103402.

\bibitem{Language1}
D.~M. Abrams, S.~H. Strogatz, Modelling the dynamics of language death, Nature
  424 (2003) 900.

\bibitem{Language4}
F.~Vazquez, X.~Castell\'o, M.~{San Miguel}, Agent-based models of language
  competition: Macroscopic descriptions and order-disorder transition, J. Stat.
  Mech. Theory Exp 2010(04) (2010) P04007.

\bibitem{Markets1}
S.~Alfarano, T.~Lux, F.~Wagner, Time estimation of agent-based models: The case
  of an asymmetric herding model, Comput. Econ 26 (2005) 19.

\bibitem{Lux2016}
T.~Lux, S.~Alfarano, Financial power laws: Empirical evidence, models, and
  mechanisms, Chaos, Solitons {\&} Fractals 88 (2016) 3--18.

\bibitem{Candia}
J.~Candia, M.~C. Gonz\'alez, P.~Wang, T.~Schoenharl, G.~Madey, A.~Barab\'asi,
  Uncovering individual and collective human dynamics from mobile phone
  records, J. Phys. A: Math. Theor 41 (2008) 224015.

\bibitem{Karsai2012}
M.~Karsai, K.~Kaski, A.-L. Barab{\'{a}}si, J.~Kert{\'{e}}sz, Universal features
  of correlated bursty behaviour, Scientific Reports 2~(1) (2012) 397.

\bibitem{Karsai2018}
M.~Karsai, H.-H. Jo, K.~Kaski, Bursty Human Dynamics, Springer International
  Publishing, 2018.

\bibitem{Baxter2011}
G.~J. Baxter, A voter model with time dependent flip rates, Journal of
  Statistical Mechanics: Theory and Experiment 2011~(09) (2011) P09005.

\bibitem{Takaguchi}
T.~Takaguchi, N.~Masuda, Voter model with non-poissonian interevent intervals,
  Phys. Rev. E 84 (2011) 036115.

\bibitem{Hoffmann2012}
T.~Hoffmann, M.~A. Porter, R.~Lambiotte, Generalized master equations for
  non-poisson dynamics on networks, Phys. Rev. E 86 (2012) 046102.

\bibitem{Min1}
B.~Min, K.~i.~Goh, I.~m.~Kim, Suppression of epidemic outbreaks with
  heavy-tailed contact dynamics, Europhys. Lett 103 (2013) 50002.

\bibitem{Boguna2014}
M.~Bogu{\~{n}}{\'{a}}, L.~Lafuerza, R.~Toral, M.~{\'{A}}. Serrano, Simulating
  non-markovian stochastic processes, Physical Review E 90 (2014) 042108.

\bibitem{Masuda2018}
N.~Masuda, L.~E.~C. Rocha, A gillespie algorithm for non-markovian stochastic
  processes, {SIAM} Review 60~(1) (2018) 95--115.

\bibitem{JEDRZEJEWSKI2018306}
A.~J\polhk{e}drzejewski, K.~Sznajd-Weron, Impact of memory on opinion dynamics,
  Physica A: Statistical Mechanics and its Applications 505 (2018) 306 -- 315.

\bibitem{Gleeson}
M.~Starnini, J.~P. Gleeson, M.~Bogu{\~n}\'a, Equivalence between non-markovian
  and {M}arkovian dynamics in epidemic spreading processes, Phys. Rev. Lett 118
  (2017) 128301.

\bibitem{Feng2019}
M.~Feng, S.-M. Cai, M.~Tang, Y.-C. Lai, Equivalence and its invalidation
  between non-markovian and markovian spreading dynamics on complex networks,
  Nature Communications 10~(1).

\bibitem{nonMarkov1}
N.~G. van Kampen, Remarks on non-markov processes, Braz. J. Phys 28 (1998) 90.

\bibitem{nonMarkov2}
J.~\L{}ucza, Non-markovian stochastic processes: Colored noise, Chaos 15 (2005)
  026107.

\bibitem{perez2016competition}
T.~P{\'e}rez, K.~Klemm, V.~M. Egu{\'\i}luz, Competition in the presence of
  aging: dominance, coexistence, and alternation between states, Scientific
  Reports 6 (2016) 21128.

\bibitem{Redner}
D.~Considine, S.~Redner, H.~Takayasu, Comment on ``{N}oise-induced bistability
  in a {M}onte {C}arlo surface-reaction model", Phys. Rev. Lett 63 (1989) 2857.

\bibitem{kirman1993ants}
A.~Kirman, Ants, rationality, and recruitment, The Quarterly Journal of
  Economics 108~(1) (1993) 137--156.

\bibitem{Granovsky}
B.~L. Granovsky, N.~Madras, The noisy voter model, Stoch. Proc. Appl 55 (1995)
  23.

\bibitem{Biancalani}
T.~Biancalani, L.~Dyson, A.~J. McKane, Noise-induced bistable states and their
  mean switching time in foraging colonies, Phys. Rev. Lett. 112 (2014) 038101.

\bibitem{Carro2}
A.~Carro, R.~Toral, M.~{San Miguel}, The noisy voter model on complex networks,
  Scientific Reports 6 (2016) 24775.

\bibitem{Peralta_pair}
A.~F. Peralta, A.~Carro, M.~{San Miguel}, R.~Toral, Stochastic pair
  approximation treatment of the noisy voter model, New Journal of Physics 20
  (2018) 103045.

\bibitem{Kononovicius}
A.~Kononovicius, V.~Gontis, Control of the socio-economic systems using herding
  interactions, Physica A 405 (2014) 80.

\bibitem{Carro1}
A.~Carro, R.~Toral, M.~{San Miguel}, {Markets, Herding and Response to External
  Information}, PloS one 10~(7) (2015) e0133287.

\bibitem{Nagi}
N.~Khalil, M.~{San Miguel}, R.~Toral, {Zealots in the mean-field noisy voter
  model}, Physical Review E 97 (2018) 0123101.

\bibitem{Nagi2}
N.~Khalil, R.~Toral, {The noisy voter model under the influence of
  contrarians}, Physica A 515 (2019) 81--92.

\bibitem{herrerias2019}
F.~Herrer\'{\i}as-Azcu\'e, T.~Galla, Consensus and diversity in multistate
  noisy voter models, Phys. Rev. E 100 (2019) 022304.

\bibitem{Vazquez2019}
F.~Vazquez, E.~S. Loscar, G.~Baglietto, Multistate voter model with imperfect
  copying, Phys. Rev. E 100 (2019) 042301.

\bibitem{Nagi3}
O.~Artime, N.~Khalil, R.~Toral, M.~{San Miguel}, {First-passage distributions
  for the one-dimensional Fokker-Planck equation}, Physical Review E 98 (2018)
  042143.

\bibitem{Schweitzer}
H.~U. Stark, C.~J. Tessone, F.~Schweitzer, Decelerating microdynamics can
  accelerate macrodynamics in the voter model, Phys. Rev. Lett 101 (2008)
  018701.

\bibitem{Juan}
J.~F. Gracia, V.~M. Egu\'iluz, M.~{San Miguel}, Update rules and interevent
  time distributions: Slow ordering versus no ordering in the voter model,
  Phys. Rev. E 84 (2011) 015103(R).

\bibitem{Oriol2}
O.~Artime, J.~F. Gracia, J.~J. Ramasco, M.~{San Miguel}, Joint effect of ageing
  and multilayer structure prevents ordering in the voter model, Sci. Rep 7
  (2017) 7166.

\bibitem{Peralta2019}
A.~F. Peralta, N.~Khalil, R.~Toral, Ordering dynamics in the voter model with
  aging, Physica A: Statistical Mechanics and its Applications (2019) 122475.

\bibitem{Oriol}
O.~Artime, A.~F. Peralta, R.~Toral, J.~J. Ramasco, M.~{San Miguel},
  Aging-induced phase transition, Phys. Rev. E 98 (2018) 032104.

\bibitem{Artime2019}
O.~Artime, A.~Carro, A.~F. Peralta, J.~J. Ramasco, M.~{San Miguel}, R.~Toral,
  Herding and idiosyncratic choices: Nonlinearity and aging-induced transitions
  in the noisy voter model, Comptes Rendus Physique 20~(4) (2019) 262--274.

\bibitem{Nyczka2012}
P.~Nyczka, K.~Sznajd-Weron, J.~Cis\l{}o, Phase transitions in the $q$-voter
  model with two types of stochastic driving, Phys. Rev. E 86 (2012) 011105.

\bibitem{Peralta}
A.~F. Peralta, A.~Carro, M.~{San Miguel}, R.~Toral, Analytical and numerical
  study of the non-linear noisy voter on complex networks, Chaos 28 (2018)
  075516.

\bibitem{Jdrzejewski2019}
A.~J{\polhk{e}}drzejewski, K.~Sznajd-Weron, Statistical physics of opinion
  formation: Is it a {SPOOF}?, Comptes Rendus Physique 20~(4) (2019) 244--261.

\bibitem{Markets2}
S.~Alfarano, T.~Lux, F.~Wagner, Time variation of higher moments in a financial
  market with heterogeneous agents: An analytical approach, Econ. Dyn. Control
  32 (2008) 101.

\bibitem{alfarano2008time}
S.~Alfarano, T.~Lux, F.~Wagner, Time variation of higher moments in a financial
  market with heterogeneous agents: An analytical approach, Journal of Economic
  Dynamics and Control 32~(1) (2008) 101--136.

\bibitem{Markets3}
S.~Alfarano, M.~Milakovi\'c, Network structure and n-dependence in agent-based
  herding models, Journal of Economic Dynamics and Control 33 (2009) 78.

\bibitem{Peralta_voter}
A.~Peralta, N.~Khalil, R.~Toral, Ordering dynamics in the voter model with
  aging., Physica A (in press).

\bibitem{Ozaita}
J.~Ozaita, Noisy voter model with partial aging and anti-aging, Master's
  thesis, University of the Balearic Islands, Palma, Spain (2018).

\bibitem{vKampen}
N.~G. van Kampen, Stochastic Processes in Physics and Chemistry, North-Holland,
  Amsterdam, 2007.

\bibitem{Peralta_moments}
A.~F. Peralta, R.~Toral, System-size expansion of the moments of a master
  equation, Chaos 28 (2018) 106303.

\bibitem{SanMiguel:1982}
J.~M. Sancho, M.~{San Miguel}, D.~D\"urr, Adiabatic elimination for systems of
  {B}rownian particles with nonconstant damping coefficients, Journal of
  Statistical Physics 28 (1982) 291.

\bibitem{Oppo:1986}
G.~L. Oppo, A.~Politi, Improved adiabatic elimination in laser equations,
  Europhysics Letters 1 (1986) 549.

\bibitem{Brion:2007}
E.~Brion, L.~H. Pedersen, K.~M{\o}lmer, Adiabatic elimination in a lambda
  system, Journal of Physics A: Mathematical and Theoretical 40 (2007) 1033.

\bibitem{pineda:2009}
M.~Pineda, R.~Toral, External noise-induced phenomena in {CO} oxidation on
  single crystal surfaces, J. Chem. Phys 130 (2009) 124704.

\bibitem{Haken}
H.~Haken, Synergetics, Springer, Berlin, 1977.

\bibitem{Lugiato}
L.~A. Lugiato, P.~Mandel, L.~M. Narducci, Adiabatic elimination in nonlinear
  dynamical systems, Phys. Rev. A 29 (1984) 1438.

\bibitem{Ariel}
A.~Fern\'andez, Center-manifold extension of the adiabatic-elimination method,
  Phys. Rev. A 32 (1985) 3070.

\bibitem{Castellano2009}
C.~Castellano, S.~Fortunato, V.~Loreto, {Statistical physics of social
  dynamics}, Rev. Mod. Phys. 81~(2) (2009) 591--646.

\bibitem{abrams2003linguistics}
D.~M. Abrams, S.~H. Strogatz, {Linguistics: Modelling the dynamics of language
  death}, Nature 424~(6951) (2003) 900.

\bibitem{Nowak1990}
A.~Nowak, J.~Szamrej, B.~Latan{\'{e}},
  \href{http://doi.apa.org/getdoi.cfm?doi=10.1037/0033-295X.97.3.362}{{From
  private attitude to public opinion: A dynamic theory of social impact.}},
  Psychological Review 97~(3) (1990) 362--376.
\newblock \href {http://dx.doi.org/10.1037/0033-295X.97.3.362}
  {\path{doi:10.1037/0033-295X.97.3.362}}.
\newline\urlprefix\url{http://doi.apa.org/getdoi.cfm?doi=10.1037/0033-295X.97.3.362}

\bibitem{touchettelarge2009}
H.~Touchette, The large deviation approach to statistical mechanics, Phys. Rep.
  478 (2009) 1.

\end{thebibliography}

\end{document}